\documentclass[preprint,12pt]{aastex}
\usepackage{rotating}
\usepackage{graphicx}

\def\eg{{e.g.,}}

\def\etal{{et~al.\null}}
\def\ie{{i.e.,}}

\def\h7{{h_{70}}}

\title{The Evolution of Ly$\alpha$ Emitting Galaxies Between 
$z = 2.1$ and $z = 3.1$}

\begin{document}
\shorttitle{Evolution of Ly$\alpha$ Emitters}

\author{Robin Ciardullo, Caryl Gronwall, Christopher Wolf,
Emily McCathran}
\affil{Department of Astronomy \& Astrophysics, The Pennsylvania State
University,  University Park, PA 16802}
\email{rbc@astro.psu.edu, caryl@astro.psu.edu, 
caw5100@psu.edu, emm5059@gmail.com}

\author{Nicholas A. Bond, Eric Gawiser}
\affil{Department of Physics and Astronomy, Rutgers, The State University 
of New Jersey, Piscataway, NJ 08854}
\email{nicholas.bond@nasa.gov, gawiser@physics.rutgers.edu}

\author{Lucia Guaita}
\affil{Departmento de Astronomia y Astrofisica, Universidad Catolica de 
Chile, Santiago, Chile}
\email{lucia.guaita@gmail.com}

\author{John J. Feldmeier}
\affil{Department of Physics \& Astronomy, Youngstown State University,
Youngstown, OH 44555-2001}
\email{jjfeldmeier@ysu.edu}

\author{Ezequiel Treister}
\affil{Institute for Astronomy, University of Hawaii, Honolulu, HI 96822}
\email{treister@ifa.hawaii.edu}

\author{Nelson Padilla, Harold Francke}
\affil{Departmento de Astronomia y Astrofisica, Universidad Catolica de 
Chile, Santiago, Chile}
\email{n.d.padilla@gmail.com, hfrancke@astro.puc.cl}

\author{Ana Matkovi\'c}
\affil{Department of Astronomy \& Astrophysics, The Pennsylvania State
University,  University Park, PA 16802}
\email{matkovic@astro.psu.edu}

\author{Martin Altmann}
\affil{Center for Astronomy, University of Heidelberg, Heidelberg,
Germany}
\email{maltmann@ari.uni-heidelberg.de}

%\and

\author{David Herrera}
\affil{NOAO, P.O. Box 26732, Tucson, AZ 85726}
\email{dherrera@noao.edu}

\begin{abstract}
We describe the results of a new, wide-field survey for $z = 3.1$ 
Ly$\alpha$ emission-line galaxies (LAEs) in the Extended Chandra Deep Field 
South (ECDF-S).  By using a nearly top-hat 5010~\AA\ filter and complementary
broadband photometry from the MUSYC survey, we identify a complete
sample of 141 objects with monochromatic fluxes brighter than
$2.4 \times 10^{-17}$~ergs~cm$^{-2}$~s$^{-1}$ and observers-frame equivalent
widths greater than $\sim 80$~\AA\ (\ie\ 20~\AA\ in the rest-frame of
Ly$\alpha$).  The bright-end of this dataset is dominated by x-ray sources
and foreground objects with GALEX detections, but when these interlopers
are removed, we are still left with a sample of 130 LAE candidates, 39 of
which have spectroscopic confirmations.  This sample overlaps the set of 
objects found in an earlier ECDF-S survey, but due to our filter's redder 
bandpass, it also includes 68 previously uncataloged sources.  We confirm 
earlier measurements of the $z=3.1$ LAE emission-line luminosity function, 
and show that an apparent anti-correlation between equivalent width
and continuum brightness is likely due to the effect of correlated 
errors in our heteroskedastic dataset.  Finally, we compare the properties
of $z = 3.1$ LAEs to LAEs found at $z = 2.1$.  We show that in the 
$\sim 1$~Gyr after $z \sim 3$, the LAE luminosity function evolved 
significantly, with $L^*$ fading by $\sim 0.4$~mag, the number density 
of sources with $L > 1.5 \times 10^{42}$~ergs~s$^{-1}$ declining by 
$\sim 50\%$, and the equivalent width scale-length contracting from 
$70^{+7}_{-5}$~\AA\ to $50^{+9}_{-6}$~\AA\null.  When combined with
literature results, our observations demonstrate that over the redshift range 
$z \sim 0$ to $z \sim 4$,  LAEs contain less than $\sim 10\%$ of the 
star-formation rate density of the universe. 

\end{abstract}


\section{Introduction}
In the local universe, relatively few galaxies have Ly$\alpha$ present in 
emission \citep{deharveng, cowie}.  This is simply due to the nature of the 
line:  since Ly$\alpha$ involves the ground state of hydrogen, each photon 
must resonantly scatter many times before escaping into intergalactic 
space.  As a result, even a small amount of dust can quench the emission 
leaving the galaxy.  

In the high-redshift universe, however, the presence of Ly$\alpha$
emission is common.  Surveys of Lyman break galaxies (LBGs) have 
demonstrated that roughly half of these bright, high-mass systems
have Ly$\alpha$ visible in emission \citep{shapley, reddy}.  Moreover, 
lower-mass Ly$\alpha$ emitting objects with extremely high mass-specific 
star-forming rates exist in large numbers \citep{gawiser06, gawiser07, 
ouchi+08}.  These Ly$\alpha$ emitters (LAEs) are an important constituent of 
the high-redshift universe.  Not only do LAEs sample a wider range of the 
galaxy luminosity function than LBGs, but they are also more 
amenable to spectroscopy.  Consequently, LAEs are powerful probes
of large scale structure, and perhaps the best tracers we have for the 
signature of baryonic acoustic oscillations.  In fact, LAEs are the targets 
of the Hobby-Eberly Telescope Dark Energy Experiment \citep{hetdex}, a 
Stage~III Dark Energy project \citep{detf} which will survey the universe 
between $1.9 < z < 3.5$.

Unfortunately, we do not yet have a clear understanding of the global
properties of LAEs.  Although their space density, luminosity function,
and equivalent width distribution are known at redshifts $z \gtrsim 3.1$, 
the evolution of these properties towards lower redshifts is uncertain.  
For instance, the results of a intermediate-bandpass (FWHM = 130~\AA) survey
with the ESO 2.2-m telescope \citep{nilsson09a} imply that LAEs become
rarer and redder by $z \sim 2.3$, but the double-blind VLT observations
of Ly$\alpha$ and H$\alpha$ emission argue otherwise \citep{hayes}.
Similarly, while the analysis of \citet{nilsson11} suggests that the mean
stellar mass of LAEs becomes greater at lower redshift, a comparison of 
narrow-band selected LAE samples at $z=3.1$ \citep{gawiser07} and $z=2.1$ 
\citet{guaita+11} show no evolution in mass.

Here we present the results of a new narrow-band survey for
$z=3.1$ LAEs in the Extended Chandra Deep Field South (ECDF-S\null).
In \S 2, we describe our observations, review the techniques used to 
detect emission-line galaxies, and give a list of LAE candidates,
130 of which are members of a statistically complete sample with
Ly$\alpha$ fluxes greater than $2.4 \times 10^{-17}$~ergs~cm$^{-2}$~s$^{-1}$ 
and rest-frame equivalent widths greater than 20~\AA\null.  In \S 3, 
we present the $z \sim 3.1$ LAE luminosity function and
show the distribution of Ly$\alpha$ equivalent widths as a function of 
line luminosity and continuum luminosity.  In \S 4, we compare these 
functions to those of the sample of $z \sim 2.1$ objects obtained by 
\citet{guaita+10}.  We show that there has been significant evolution 
between the two epochs in $L^*$, the characteristic emission-line 
luminosity of the \citet{schechter} function, $\phi$, the integrated number
density of bright LAEs, and $w_0$, the e-folding scale 
length of the rest-frame equivalent width distribution.  In \S 5,
we examine the LAEs' star-formation rates, using both their rest-frame
UV emission and Ly$\alpha$, and in \S 6, we compare the star-formation
rate density of the LAE population with that of the universe as a whole.
We conclude by summarizing our results and arguing that LAEs represent the
extreme low-metallicity, low-extinction, faint-end tail of the high-redshift
star-forming galaxy population.

For this paper, we assume a $\Lambda$CDM cosmology, with 
$\Omega_{\Lambda} = 0.7$, $\Omega_M = 0.3$ and 
$H_0 = 70$~km~s$^{-1}$~Mpc$^{-1}$.

\section{The $z=3.1$ Sample}
\begin{figure}[t]
\figurenum{1}
\plotone{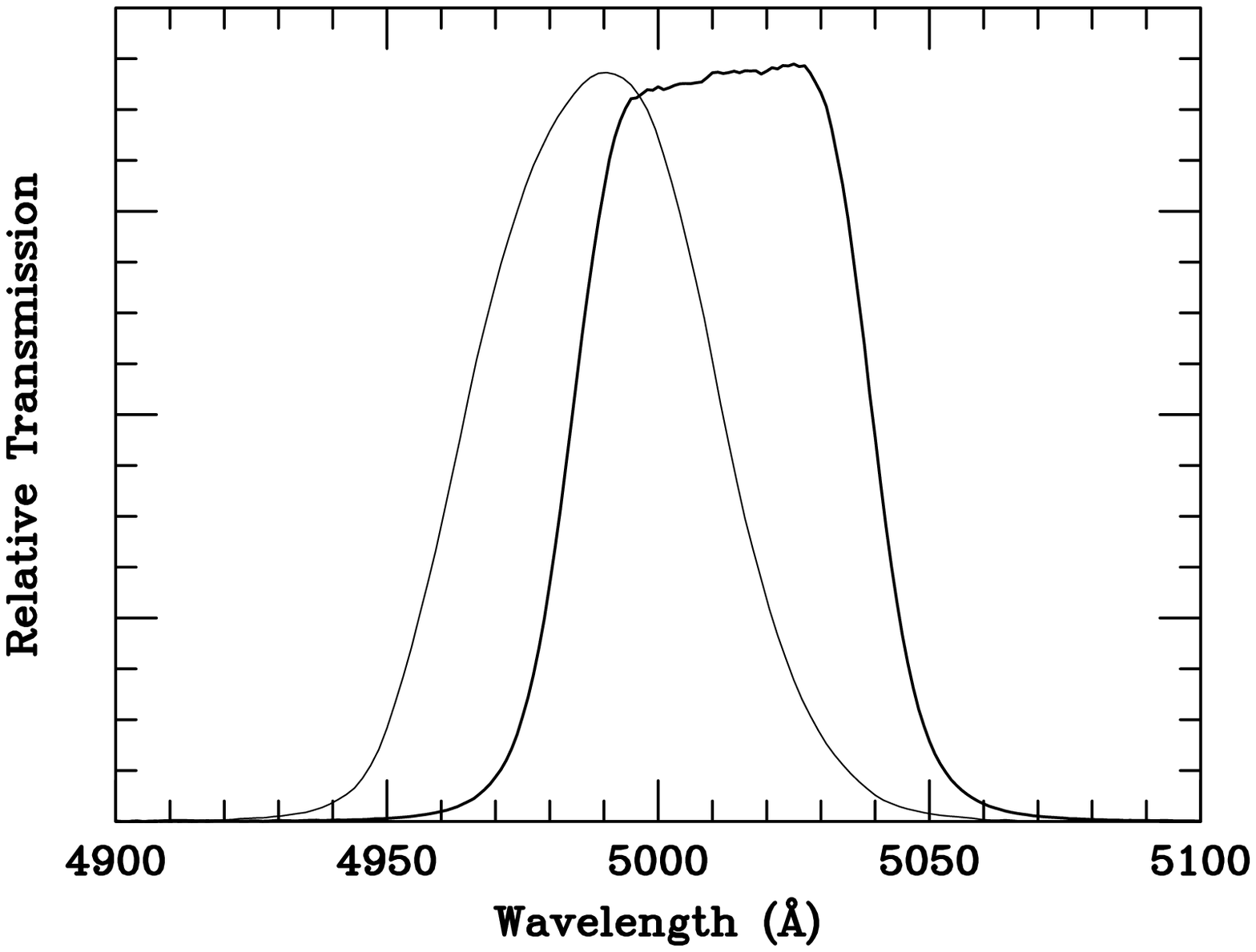}
\caption[o3_filters]{Transmission curve of our narrow-band $\lambda 5010$ 
filter in the f/2.7 beam of the CTIO 4-m telescope, compared to that of the 
4990~\AA\ filter used by \citet{g+07}.  Our filter's nearly top-hat profile 
simplifies the analysis of the current survey.}
\label{o3_filters}
\end{figure}

The luminosity function and equivalent width distribution of $z=3.1$
LAEs is fairly well-known from the narrow-band surveys of
\citet[][hereafter Gr07]{g+07} and \citet{ouchi+08}.  However these 
observations, which have identified more than 200 LAE candidates with 
little contamination from foreground objects, suffer from larger than 
normal photometric uncertainties due to the Gaussian-like transmission 
profiles of their narrow-band filters.  To confirm these results, and 
to increase the survey sample at $z \sim 3.1$, we extended the Gr07 study 
by re-imaging the ECDF-S, this time with a 57~\AA\ full-width-half-maximum 
(FWHM) nearly top-hat filter centered near 5010~\AA\null.  As with Gr07, 
the data were collected with the MOSAIC~II CCD camera on the CTIO Blanco 
4-m telescope, and consisted of a set of 47 images taken in $1\farcs 1$ 
seeing for a total narrow-band exposure time of 15.67~hr.  A log of the 
observations is given in Table~\ref{obslog}; a figure showing our filter's 
transmission curve, compared to that used by Gr07, appears in 
Figure~\ref{o3_filters}.

\begin{figure}[t]
\figurenum{2}
\plotone{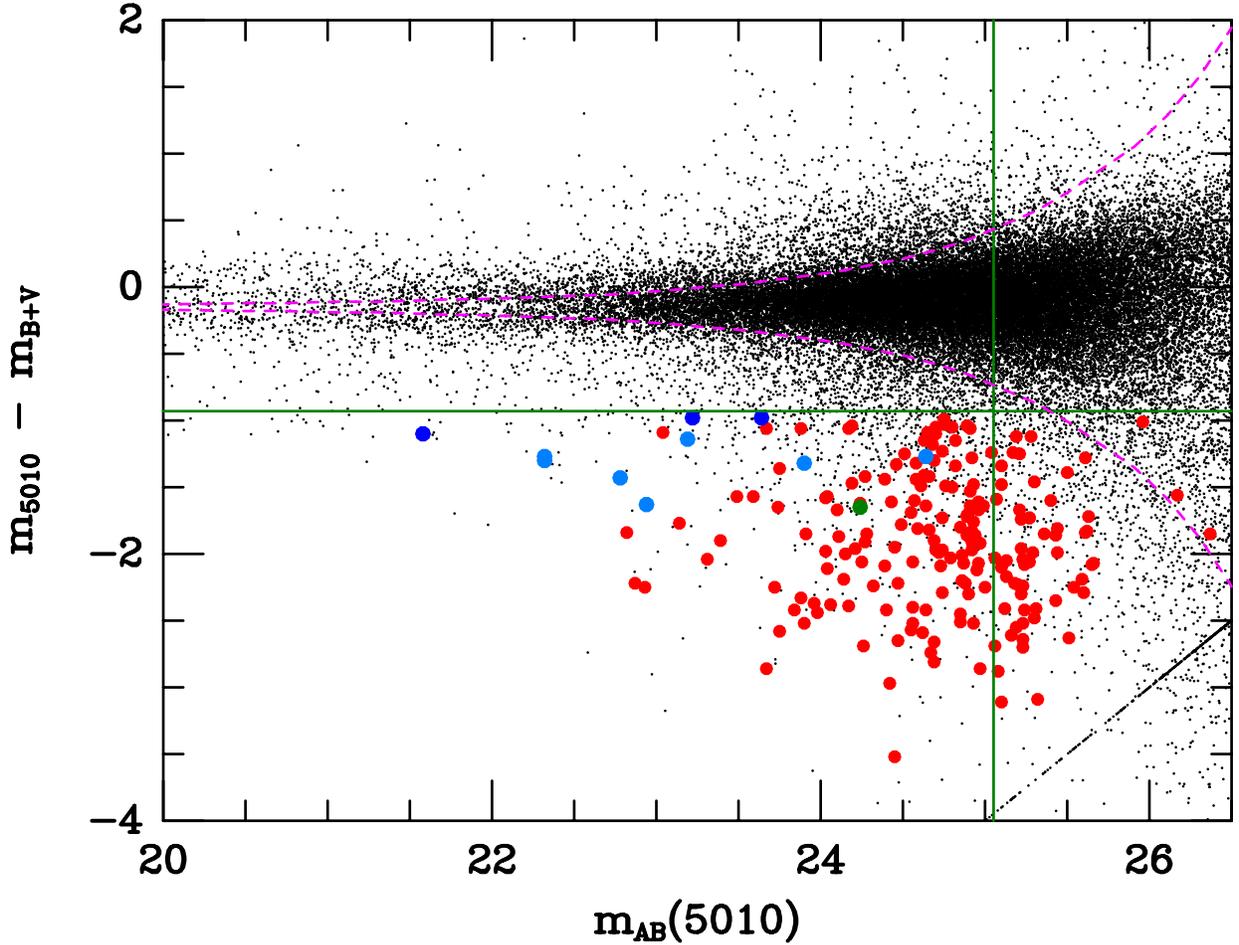}
\caption[cmd]{Excess emission in the narrow-band $\lambda 5010$ filter over 
the continuum for objects in our survey field.  The abscissa gives $\lambda 
5010$ AB magnitude, while the ordinate shows the difference between this 
magnitude and the magnitude on a $B+V$ continuum frame.  Our narrow-band 
completeness limit of $2.4 \times 10^{-17}$~ergs~cm$^{-2}$~s$^{-1}$ is 
represented by a vertical line; our equivalent width limit of 80~\AA\ is shown 
via the horizontal line at $m_{5010} - m_{B+V} = -0.93$.  The magenta curve 
shows the expected $3 \, \sigma$ errors in the photometry.  Candidate emission 
line galaxies are denoted as red circles; the green circle shows an x-ray 
source detected by Chandra, the dark blue circles denote ultraviolet sources 
cataloged by GALEX and/or Swift/UVOT, and the light blue 
circles show objects detected in both the x-ray (with Chandra) and UV 
(with GALEX and/or Swift/UVOT).}
\label{cmd}
\end{figure}

The procedures used to reduce the data, identify line emitters, and measure
LAE brightnesses were identical to those used by Gr07 and discussed in
detail by \citet{ipn2}.  After de-biasing, flat-fielding, and 
aligning the data, we co-added our narrow-band frames to create
a master image that was clipped of cosmic rays.  This frame was then
compared to a deep $B$+$V$ continuum image provided by the MUSYC 
collaboration \citep{musyc}, and the routines of DAOPHOT were 
used to create a color-magnitude diagram of all sources found
on the narrow-band image.  Objects with narrow-band minus continuum colors 
less than 0.93~mag in the AB system were flagged as possible emission-line 
sources (see Figure~\ref{cmd}).   Then, to detect those LAEs whose photometry
was compromised by object blending, sky gradients, and/or source 
confusion, we re-ran our detection algorithm on a ``difference'' frame made 
by subtracting a scaled version of the $B$+$V$ continuum image from the 
narrow-band frame.  In this case, our detection algorithm was set to flag 
all objects brighter than four times the standard deviation of the local
sky.  This difference technique produced a sample of several
thousand additional candidates, most of which were below our equivalent
width threshold.  However, when the weak emission-line sources were
excluded, the result was a sample of objects that was $\sim 15\%$ larger
than that produced by the two-color method alone.

For both detection algorithms, we intentionally biased our parameters to 
identify faint sources at the expense of false detections (\ie\ we introduced 
Type~I errors into our dataset).  To compensate for this, we 
visually inspected each candidate, both on the co-added narrow-band
and broadband frames, and on sub-samples of the original images.  
This step excluded most of the false positives, which were typically at 
the frame limit, and artifacts produced by cross-talk associated with the 
CCD electronics.

To derive the line fluxes and equivalent widths of our LAE candidates,
we took advantage of the fact that Gr07 measured the 5000~\AA\ AB 
magnitudes of several ECDF-S field stars by comparing their large aperture 
instrumental magnitudes to those of spectrophotometric standard stars
\citep{stone, hamuy92} taken throughout their survey.  To place our new 
5010~\AA\ narrow-band observations on the same photometric system, we 
therefore measured our LAE brightnesses with respect to these same stars.  We 
then converted our narrow-band 5010 AB magnitudes into monochromatic 
fluxes ($F_{5010}$) by comparing the filter's integral transmission (which 
is the relevant quantity for field star photometry) to its monochromatic 
transmission at the center of the bandpass \citep[see Gr07 and][]{jqa}.  Note 
that this simple procedure is only exact for top-hat filters where the 
monochromatic transmission is insensitive to wavelength:  as Gr07 showed, 
observations through filters with a Gaussian-shaped transmission curve require 
a more sophisticated approach.  

Along with monochromatic flux, we also derived photometric equivalent
widths for all our LAE candidates by comparing their narrow-band AB flux 
densities, $f_{5010}$, to their $B+V$ continuum flux densities, 
$f_{B+V}$ via
\begin{equation}
{\rm EW} = \left\{ {f_{5010} \over f_{B+V}} - 1 \right\} \Delta \lambda
\label{eq_ew}
\end{equation}
where $\Delta\lambda = 57$~\AA\ approximates the contribution of the galaxy's
underlying continuum within the narrow-band filter's bandpass.
For consistency with Gr07, we then excluded all sources with
equivalent widths less than 20~\AA\ in the rest-frame of Ly$\alpha$
(approximately 80~\AA\ in the observers frame), and derived equatorial
positions for the remaining objects with respect to reference stars in
the USNO-B1.0 astrometric catalog \citep{usno-b}.  This left us with a sample 
of 199 LAE candidates.

To measure the completeness of our sample as a function of
magnitude, we followed the procedures described by \citet{ipn2} and
performed a series of artificial star experiments.  By adding 
1,000,000 stars (2000 at a time) to our narrow-band frame, and
re-running our detection algorithms, we were able to compute the flux level 
below which the object recovery fraction dropped below 90\% and 50\%.  The
90\% value, which corresponds to an AB magnitude of $m_{5010} \sim 25.05$, 
and a monochromatic flux of $2.4 \times 10^{-17}$~ergs~cm$^{-2}$~s$^{-1}$ 
($\log F_{5010} = -16.61$) is our limiting flux for statistical completeness.  
A total of 141 objects are brighter than this limit.  Those emission-line 
objects satisfying our 80~\AA\ equivalent width threshold and brighter than
our completeness limit are listed in Table~\ref{brightLAEs}; additional
objects below the completeness limit are given in Table~\ref{faintLAEs}.
Our estimated photometric errors, as a function of log flux, appear in 
Table~\ref{errors}. 

\begin{figure}[t]
\figurenum{3}
\plotone{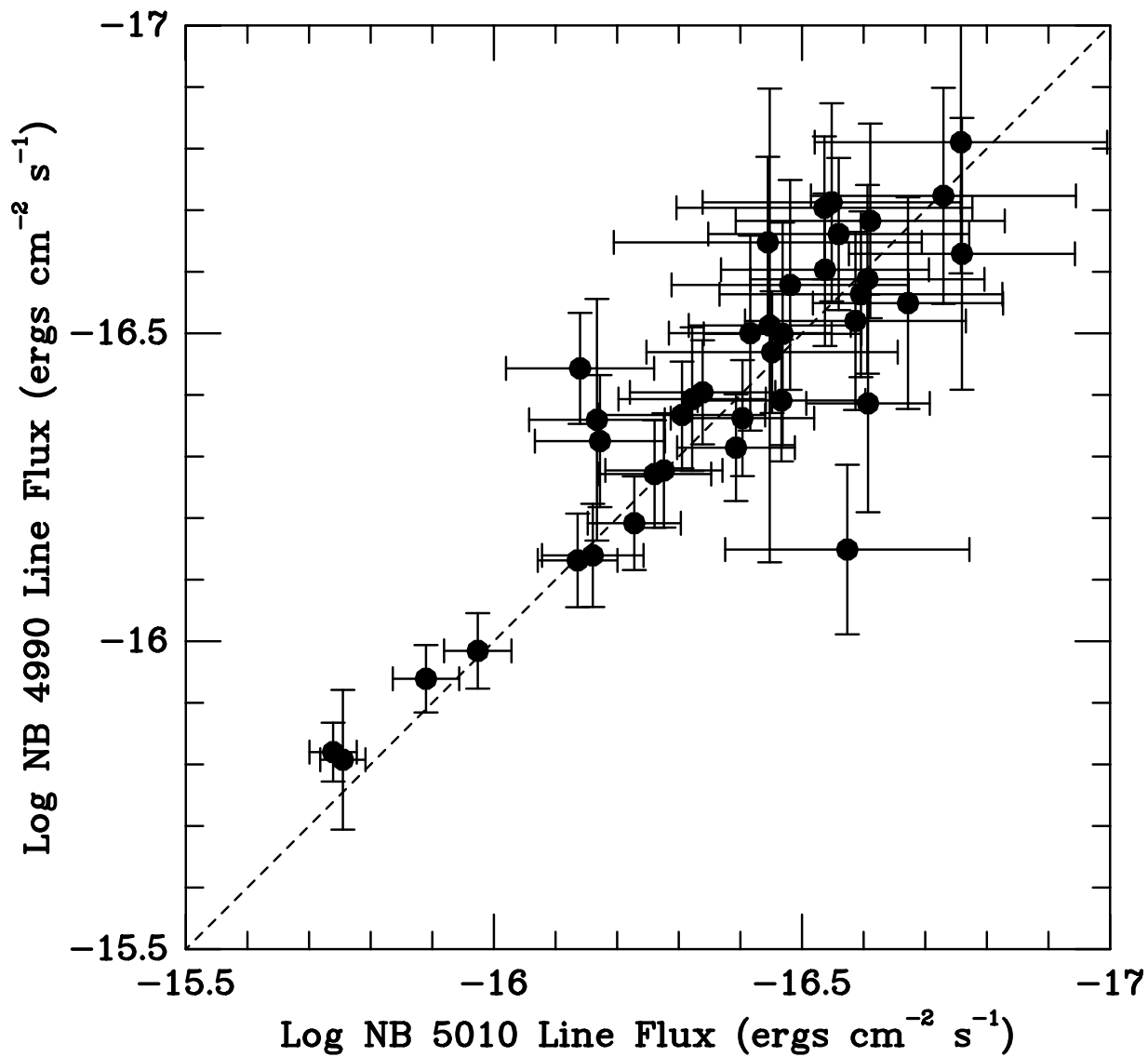}
\caption[flux_compare]{A comparison of the continuum-subtracted emission-line 
fluxes of 39 spectroscopically confirmed LAEs detected with both a narrow-band 
$\lambda 4990$ filter and our narrow-band $\lambda 5010$ filter. 
The scatter is consistent with the photometric precision 
of the measurements.}
\label{flux_compare}
\end{figure}

To confirm the robustness of our measurements, and to double-check the 
transmission curve of our narrow-band filter, we compared our sample of 
Ly$\alpha$ emitters to the list of LAEs published by Gr07\null.  As 
Figure~\ref{o3_filters} shows, the volumes studied in these two 
surveys have a great deal of overlap:  a quick inspection of the two curves
suggests that $\sim 40\%$ of the LAEs found in our new survey should appear 
in the Gr07 tables.    Indeed, this is the case: of the 199 candidates 
detected, 80 are present in the previous list of $z \sim 3.1$ emission-line 
sources.  Moreover, 39 of the objects common to both surveys already have 
spectroscopic confirmations.  In this restricted sample, the lowest redshift 
object has its emission line at 4981~\AA\ ($z=3.096$), where the transmission 
of our filter falls to roughly 25\% of its peak value.  Conversely, our 
highest redshift source has Ly$\alpha$ at 5023~\AA\ ($z = 3.131$), \ie\ 
where the Gr07 filter transmission is $\sim 20\%$ of its peak.  A comparison 
of the inferred monochromatic fluxes for this subset of objects, where the 
wavelength of the Ly$\alpha$ emission-line is known, also shows excellent 
agreement (see Figure~\ref{flux_compare}).  This consistency argues that the
transmission curves displayed in Figure~\ref{o3_filters} are accurate, and
the photometric errors tabulated in Table~\ref{errors} are reasonable.

\subsection{Chandra, GALEX, and Swift Detections}

\begin{figure}[t]
\figurenum{4}
\plotone{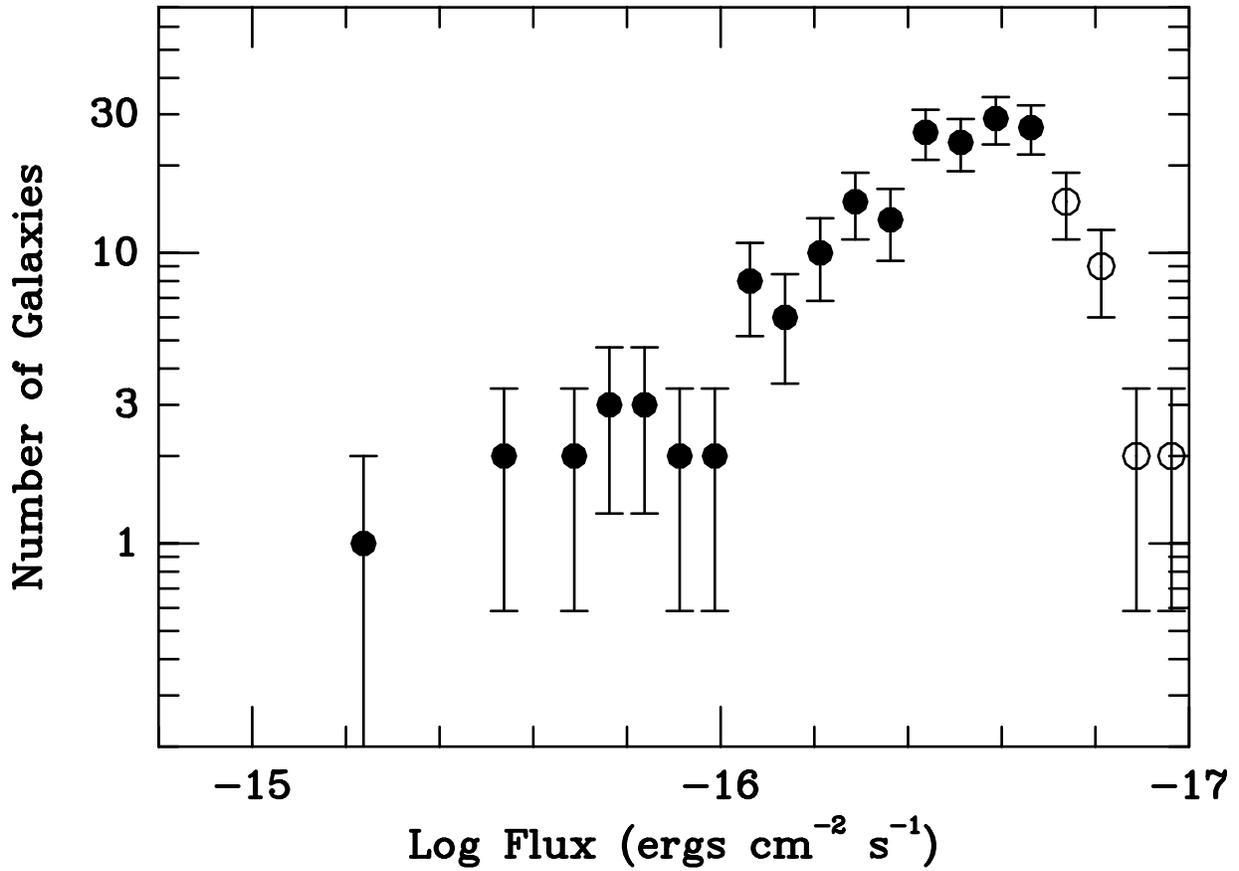}
\caption[lf_raw]{The monochromatic flux distribution of objects with observed 
equivalent widths EW $> 80$~\AA.  The brightest source is a resolved
[O~II] emitter; the next three brightest objects are x-ray sources.
The open circles represent data beyond our completeness limit.}
\label{lf_raw}
\end{figure}

Figure~\ref{lf_raw} shows the flux distribution of all our high-equivalent
width (EW $> 80$~\AA) objects.  The most obvious feature is the presence of 
a small number of sources with monochromatic fluxes brighter than 
$\sim 10^{-16}$~ergs~cm$^{-2}$~s$^{-1}$.  This luminous tail is not nearly
as pronounced in the $z=3.1$ survey of \citet{ouchi+08}, and not
visible at all in the Gr07 dataset.  A closer examination, however, reveals
that many of these extremely bright sources are foreground interlopers.

To show this, we first removed the AGN from our sample by cross-correlating
the positions of our LAE candidates with those of the x-ray point sources 
found in the 2~Msec exposure of the Chandra Deep Field South \citep{luo}
and the four 250~ksec exposures of the Extended Chandra Deep Field South
\citep{lehmer05, virani}.  Eight spatial coincidences were found, including 
one for our second brightest source \citep[a $z = 1.609$ AGN which entered our 
sample due to its strong C~III{]} $\lambda 1909$ emission;][]{balestra}, and 
our third brightest source \citep[which, at $z=0.977$, must have been detected 
due to a change in brightness between the time of our narrow-band and continuum
exposures;][]{szokoly}.  In fact, at flux levels brighter than 
$\log F > -15.9$, half of our LAE candidates (and the candidates of Gr07) are 
co-incident with x-ray identified AGN\null.  These contaminants are noted in
Table~\ref{brightLAEs}.

We next compared our list of LAE candidates to the catalogs of UV sources
detected by the GALEX \citep{Galex} and Swift \citep[UVOT;][]{UVOT}
satellites.  Both instruments have conducted deep ($m_{AB} \sim 25$)
surveys of the ECDF-S using near-UV (and, for GALEX, far-UV) filters.
Since at $z \sim 3.1$, the near-UV corresponds to rest-frame wavelengths far
beyond the Lyman break, any LAE candidate listed in these catalogs
is most likely a foreground contaminant.  Our cross-correlation procedure 
(which used a matching radius of $2\arcsec$), produced ten coincidences,
including five associated with the x-ray sources described above.  Of the 
remaining objects, the most prominent is our LAE candidate \#1, 
a marginally-resolved galaxy which is nearly a magnitude brighter than any 
other object in our sample.  Fortuitously, this object already has a redshift:  
spectroscopy by \citet{vanzella+08} confirms the source to be a 
$z = 0.337$ galaxy with extremely strong [O~II] $\lambda 3727$ emission
(rest-frame equivalent width EW$_0 \sim 73$~\AA).  Although 
such objects are rare, and none were found in the Gr07 survey,
a few are present in the magnitude-limited set of objects defined by
\citet{hogg98}.  The other four UV matches include two likely superpositions
(\ie\ objects containing a more-likely counterpart within the error circle), 
and two sources with estimated observers-frame equivalent widths (82~\AA) 
barely above our threshold.  (The spectroscopy of \citet{vvds} confirms that 
one of them is foreground.)  We identify each of these UV-bright sources in 
Table~\ref{brightLAEs}, but in the analysis that follows, we exclude 
only the x-ray sources and the three GALEX/Swift-UVOT objects with 
unambiguously identified counterparts.  This leaves us with a statistically
complete sample of 130 objects, 39 of which have spectroscopic confirmations.

\subsection{X-ray Stacking}

By cross-correlating our LAE positions with the x-ray catalog from Chandra,
we were able to eliminate bright AGN from our candidate list.  This,
of course, does not exclude the possibility that many of our LAEs have 
low-level nuclear activity.  To place a constraint on this activity, we 
performed a stacking analysis on the x-ray data, similar to that described 
by \citet{brandt}.  We combined our new LAE dataset with that found by
Gr07, excluded the known x-ray sources, and stacked the x-ray signal
of all 233 LAEs in the statistically complete samples, using the new
4~Msec exposure of the Chandra Deep Field South and the 250~ksec
exposures of the surrounding region.  The mean LAE of the resulting
$\sim 480$~Msec effective exposure has an x-ray flux less than
$7.24 \times 10^{-8}$~counts~s$^{-1}$ in the soft band and
$1.23 \times 10^{-7}$~counts~s$^{-1}$ in the hard band ($1 \, \sigma$
upper limits), implying luminosity limits of 
$L_X < 5.3 \times 10^{40}$~ergs~s$^{-1}$ and 
$L_X < 2.5 \times 10^{41}$~ergs~s$^{-1}$, respectively.   Clearly,
few, if any of our LAEs harbor low-luminosity AGN\null.  We 
also used these data to place a limit on the mean star-formation 
rate of the LAEs in our sample.  Since $z \sim 3.1$ LAEs all have
extremely high mass-specific star-formation rates \citep{gawiser07},
the contributions of low-mass x-ray binaries to the total 2 -- 10~keV
emission of these systems should be negligible.  This being the
case, our upper limit on the hard x-ray flux from the LAE
population yields a star-formation rate of $< 33 \, M_{\odot}$~yr$^{-1}$
or $< 53 \, M_{\odot}$~yr$^{-1}$, depending on whether we use 
\citet{lehmer10} or \citet{persic} for the calibration.
Although this value is still several times larger than the star-formation
rates derived from the LAE's rest-frame UV and Ly$\alpha$ emission
(see \S 5), the numbers do provide hope that with additional observations,
we will be able to detect the x-ray emission from $z \sim 3.1$
LAE starbursts.  

\section{The $z = 3.1$ LAE Luminosity Function and Equivalent Width
Distribution}

\begin{figure}[t]
\figurenum{5}
\plotone{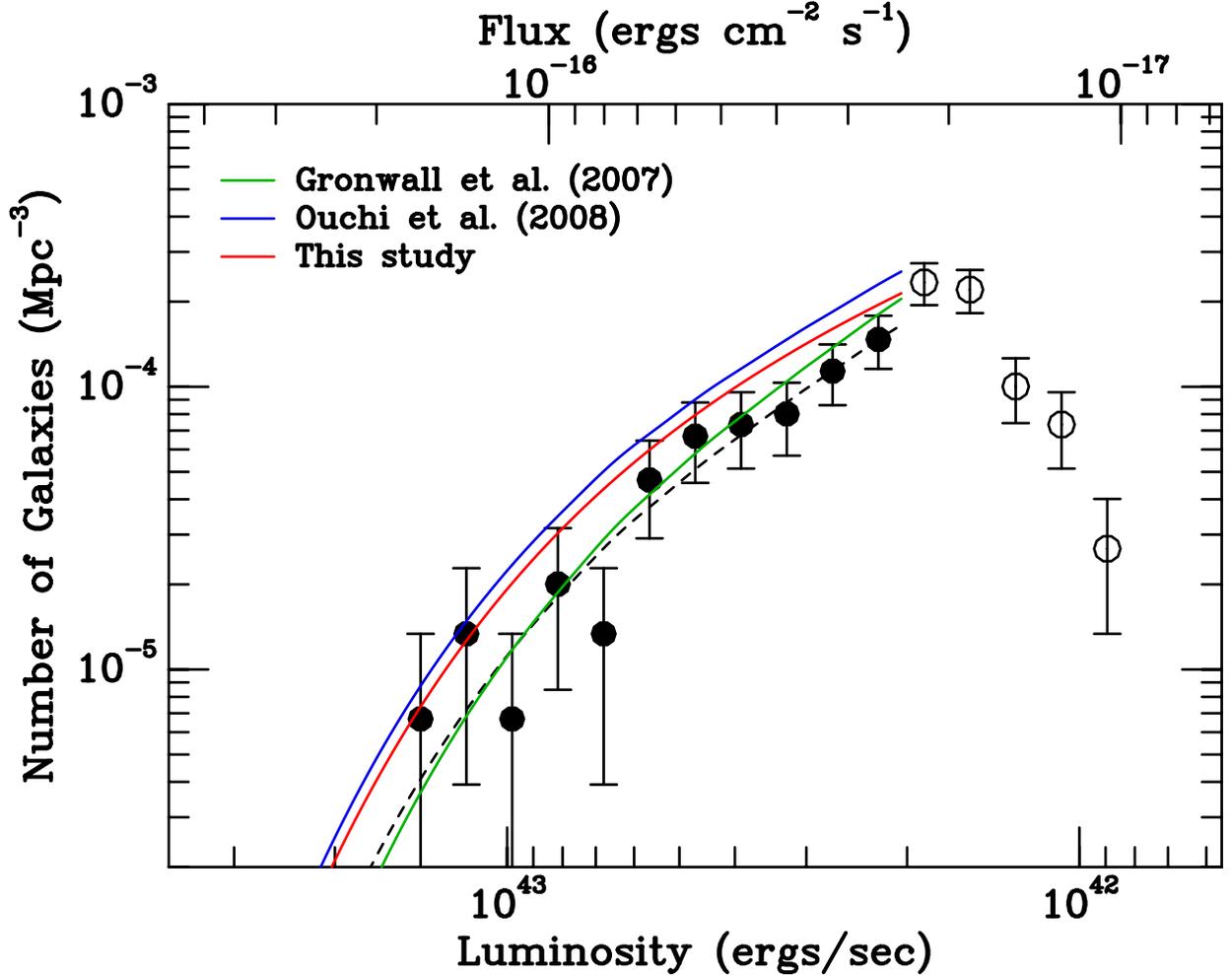}
\caption[lumfun_z3]{The differential luminosity function of $z=3.1$ Ly$\alpha$ 
galaxies with rest frame equivalent widths greater than 20~\AA\ binned into 
0.2~mag intervals.  The points give the density of objects under the 
assumption that our filter's FWHM defines the survey volume; the open circles 
represent data beyond our completeness limit.  The solid red curve shows our 
input best-fit \citet{schechter} luminosity function, while the dashed line 
illustrates the shape and normalization of this function after correcting for 
the effects of photometric error and censoring by our filter's non-square 
transmission curve.  The luminosity functions of Gr07 and \citet{ouchi+08} are 
also plotted, though the curve for the latter is based on a rest-frame 
equivalent width limit of 64~\AA\null.  For consistency with our selection 
criteria, this curve should be increased by $\gtrsim 10\%$.}
\label{lumfun_z3}
\end{figure}

Figure~\ref{lumfun_z3} displays the distribution of continuum-subtracted 
Ly$\alpha$ emission-line luminosities for our new sample of $z=3.1$ LAEs.
To find the best-fit \citet{schechter} function to these data, we followed 
the procedures described in detail by Gr07\null.  First, we computed two 
convolution kernels, one (a Gaussian) representing the photometric uncertainty 
of our measurements (as a function of magnitude), and the other determined 
using the derivative of the inverse of the filter transmission function.  
(This latter kernel does not normalize to one, since it also includes the 
effect of equivalent width censorship.)  We then applied these two kernels 
to a series of functions of the form
\begin{equation}
\phi(L/L^*) d(L/L^*) = \phi^* \left(L/L^*\right)^\alpha e^{-L/L^*} d(L/L^*),
\label{eq_schechter}
\end{equation}
treated each curve as a probability distribution, and computed the likelihood
that the observed sample of Ly$\alpha$ luminosities could be drawn from 
that distribution.  Figure~\ref{lumfun_z3} displays our best-fit function,
both before (the dashed line) and after the equivalent-width censorship
(solid red line).  Also plotted are the best-fit curves for the 
$z=3.1$ LAEs found by Gr07 (in green) and \citet[][in blue]{ouchi+08}.  
The parameters of these functions are listed in Table~\ref{sch_lf}.
Note that for the current study, we have fixed the faint-end slope of our
\citet{schechter} function to $\alpha = -1.65$.  This is the most-likely value
inferred from the Gr07 data, once continuum subtraction has been applied
to their fluxes, and it is also consistent with the recent 
$2 < z < 6.6$ LAE measurements of \citet{cassata}.  Since our data are 
$\sim 0.5$~mag shallower than that of Gr07, we adopt their estimate for 
the faint-end slope.

From the figure, it is immediately obvious that all three luminosity
functions are similar, both in shape and in normalization, and all fit
the data extremely well.  Our most-likely value of $\log L^* = 42.76 \pm
0.10$ is similar to that found by \citet{ouchi+08} and less than $\sim 0.1$~mag 
brighter than that derived Gr07\null.  Similarly, all three functions imply
$\sim 7 \times 10^{-4}$~Mpc$^{-3}$ for the density of galaxies with
rest-frame equivalent widths greater than 20~\AA\ and Ly$\alpha$ 
fluxes greater than $2.4 \times 10^{-17}$~ergs~cm$^{-2}$~s$^{-1}$.  
Since our new dataset has substantial overlap with that of Gr07,
the consistency between these two samples is not surprising.
The \citet{ouchi+08} survey, however, is an independent measurement.
Even after we follow their recommendation and increase $\phi^*$
by $\sim 10\%$ to compensate for their higher equivalent width limit,
there is still excellent agreement in the functions.  This
is indicative of the robustness of the results.

\subsection{Equivalent Width versus Luminosity}

\begin{figure}[t]
\figurenum{6}
\plotone{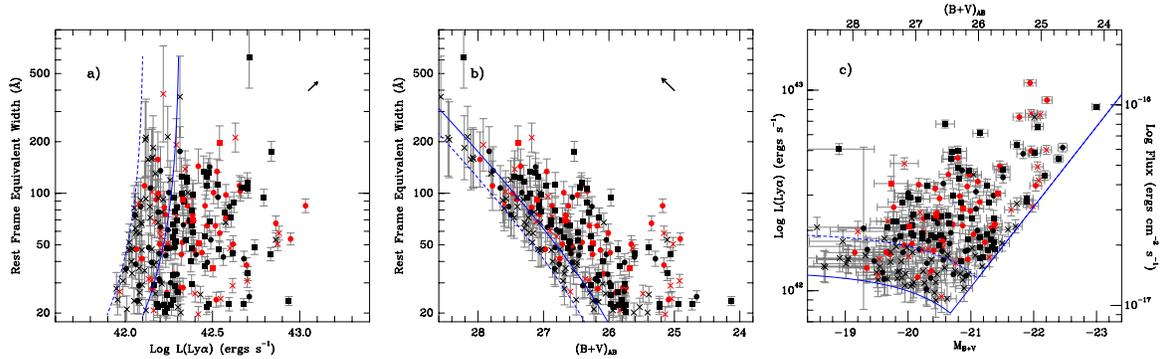}
\vskip-2.9in
\caption[ew_3part]{The left-hand panel displays the distribution of Ly$\alpha$ 
rest-frame equivalent widths as a function of (continuum subtracted) Ly$\alpha$ 
emission-line luminosity, the middle panel shows these same equivalent
widths as a function of continuum magnitude, and the right-hand
panel plots Ly$\alpha$ luminosity versus continuum luminosity.
Crosses are values from Gr07, squares are from the current study, and 
circles represent LAE candidates found in both surveys.  The red points 
show spectroscopically confirmed objects.  The lines show the completeness 
limits of the surveys (solid for this survey, dashed for Gr07), 
and the arrows in the left and middle panels show how the errors
in equivalent width and luminosity are expected to correlate.
The heteroskedastic nature of the uncertainties makes a 
simple interpretation of the equivalent width diagrams difficult,
but the last panel strongly suggests that there is no correlation
between any of the quantities.}
\label{ew_3part}
\end{figure}

There has been some discussion in the literature concerning the
relationship between Ly$\alpha$ equivalent width and continuum luminosity
in high-redshift galaxies.  Starting with \citet{ando+06}, a number of
authors have claimed a deficit of continuum bright, high-equivalent 
width objects, both in samples of Lyman-break galaxies \citep[\eg][]{ando+07,
iwata, vanzella+09} and Ly$\alpha$ emitters \citep{ouchi+08, shioya}.  
However, \citet{nilsson09b}, via a series of Monte Carlo simulations,
have argued that the observed correlations are not statistically
significant, and there is no need to invoke the presence of
exotic stellar populations or a clumpy interstellar medium
\citep{ouchi+08, mao+07, kobayashi+10}.  We illustrate the problem 
in Figure~\ref{ew_3part}.  The left panel of the figure shows the
LAE equivalent width measurements as a function of (continuum 
subtracted) Ly$\alpha$ luminosity, for both the present sample and the
sample of objects found by Gr07\null.  The figure displays no evidence for a 
shift in the equivalent width distribution as a function of luminosity.
A 2-sample Anderson-Darling test \citep{adtest} which compares the
equivalent widths of bright ($L > 3 \times 10^{42}$~ergs~s$^{-1}$) and 
faint ($1.25 \times 10^{42} < L < 3 \times 10^{42}$~ergs~s$^{-1}$) LAEs 
confirms that there is no significant difference between the two samples.  
Moreover, if we apply the non-parametric rank-order \citet{efron} test to the
data, we find no reason to reject the hypothesis that equivalent
width and Ly$\alpha$ luminosity are statistically independent quantities.

In contrast, the middle panel of Figure~\ref{ew_3part} displays the 
distribution of Ly$\alpha$ equivalent widths as a function of continuum 
magnitude.   The difference between the two diagrams is striking.  While
the lack of low equivalent-width sources with faint continuum magnitudes
is simply a selection effect (the Ly$\alpha$ emission-line is too faint 
to be seen), the absence of high equivalent-width, bright
continuum objects is real.   If we apply the non-parametric rank-order 
statistic described by \citet{efron} to our truncated dataset, then we 
find that the null hypothesis of independence between equivalent width and 
continuum magnitude is excluded at the $5 \sigma$ level (for the
Gr07 sample) and the $7 \sigma$ level (for our new $z = 3.1$ 
dataset).  In other words, equivalent width shows no correlation
when plotted against line brightness, but a clear anti-correlation when
compared to continuum brightness.

The explanation for these seemingly contradictory results probably lies 
in the fact that equivalent width is a derived quantity formed from
two photometric measurements, one for the emission line, and one
for the continuum.  Consequently, the points displayed in the first two panels
of Figure~\ref{ew_3part} possess correlated errors.  In the left-hand diagram, 
emission-line strength and equivalent width are positively correlated, so 
photometric errors will preferentially scatter objects into the 
high-luminosity, high-equivalent width region of the diagram.  Conversely, 
in the middle panel, continuum strength and equivalent width are 
negatively correlated, so errors in the abscissa will move objects
away from this portion of the figure.  When this dichotomy is coupled with 
the heteroskedastic nature of the dataset -- the largest uncertainties are
associated with the objects having the faintest continuum magnitudes and
therefore the highest equivalent widths -- the result can be an apparent 
correlation in one figure, and a lack of correlation in another.

The best way to investigate the systematics of equivalent width is to
plot the two independent quantities in the relation -- line strength
and continuum strength -- against one another.  This is done in the 
right-hand panel of Figure~\ref{ew_3part}.  Here, there is no evidence of 
any systematic behavior with luminosity:  the LAEs occupy a strip in
the diagram whose thickness appears constant with luminosity.  This
argues strongly against the presence of a systematic dependence of
equivalent width on luminosity.

\section{Evolution from $z \sim 3$ to $z \sim 2$}
To investigate the evolution of Ly$\alpha$ emitters in the redshift range
$2.1 < z < 3.1$, we can compare the properties of our LAE candidates to those
of the $z \sim 2.1$ objects found by \citet[][hereafter Gu10]{guaita+10} in 
their deep narrow-band survey of the ECDF-S\null.  Like the observations 
described here, Gu10 used a 50~\AA\ wide FWHM filter ($\lambda_c = 3727$) and 
the prime focus Mosaic~II CCD camera of the CTIO 4-m telescope to identify a
complete sample of candidate LAEs over $\sim 1000$~arcmin$^2$ region
of the sky.  Also, like our $z \sim 3.1$ sample, the Gu10 objects 
are defined via their color with respect to deep continuum images provided by 
the MUSYC survey \citep{musyc}, have rest-frame Ly$\alpha$ equivalent widths 
greater than 20~\AA, and have minimum Ly$\alpha$ luminosities of 
$\sim 1.3 \times 10^{42}$~ergs~s$^{-1}$.  Thus, in
area, number, and depth, the Gu10 dataset is well-suited for a
comparative study.

Gu10 have already used the Chandra and GALEX source lists
to eliminate some high-equivalent width interlopers from their sample:
as with our $z \sim 3.1$ candidates, true $z \sim 2.1$
galaxies without AGN should not only be undetected in the x-ray, but
in the ultraviolet as well, as nearly all their near-UV flux originates
at wavelengths beyond the Lyman break.   To these criteria, we added
two others.  First, we eliminated one LAE candidate that was previously 
identified as a $z=2.0$ QSO in the COMBO-17 source catalog \citep{combo17}.
EIS J033330.60$-$274819.3 is located at the extreme edge of the ECDF-S Field~1
x-ray pointing, $\sim 9\arcmin$ from the field center.  Although the
object is not cataloged as an x-ray source, it is (barely) detected in the
hard x-ray band, and its implied (rest-frame) 2~keV to 2500~\AA\ flux ratio 
is consistent with that expected for an AGN \citep{strateva}.  (Using a 
$1\farcs 5$ aperture, we derive $\alpha_{\rm ox} = -1.5^{+0.2}_{-0.4}$.)  
Moreover, the candidate in question is exceptionally luminous for an LAE, as
it has a continuum-subtracted narrow-band flux that is over a magnitude
brighter than the next most luminous source in the Gu10 object list.
Thus, even if the candidate is not an AGN, its extreme luminosity places
it in a different category from all other candidates in the Gu10 sample,
and unfittable via a traditional \citet{schechter} function.

In addition to excluding this one object, we also examined each candidates'
appearance on {\sl HST\/} $V$-band images taken as part of the GOODS 
\citep{GOODS} and GEMS \citep{GEMS} surveys.  Since LAEs are, in general, 
much fainter in the continuum than Lyman-break galaxies, we might expect the 
sample of $z \sim 2.1$ objects to be smaller (in physical extent) than the 
largest Lyman-break galaxies found by \citet{ferguson04} and \citet{rav06}.   
In fact, the {\sl HST\/}-based morphological study of \citet{bond11}
classified these ``large'' objects as photometric interlopers, and 
\citet{guaita+11} used this discriminant in their analysis of the LAEs'
spectral energy distributions.  Thirty-four of the Gu10 objects fall
into this category, but only four are among the brightest 100 objects in
the sample, and only one is more luminous than the sample's 90\% completeness 
limit.  Consequently, their inclusion or exclusion makes very little 
difference to the analysis.  Below, we will present results both including 
and excluding these objects.

Before fitting the luminosity function, we made one other modification to the
Gu10 dataset.   For consistency with our $z \sim 3.1$ photometry and with the
analysis of Gr07, we chose to adopt Gu10's point-source aperture fluxes
in our analysis, rather than their corrected aperture fluxes, which were
derived using the photometric techniques of \citet{musyc}.  While this latter
procedure is most useful for selecting LAEs out of the general galaxy 
population, the former method is better suited for measuring the brightnesses 
of exceedingly compact ($<0\farcs 25$) sources on groundbased images.  
(Since the seeing on the Gu10 frames is roughly twice the size of the 
largest LAE counterpart identified on {\sl HST\/} frames of the region
\citep{bond11}, this change is certainly justified.)  This technical revision 
reduces the derived magnitudes of a significant fraction of the Gu10 LAEs, 
sometimes by as much as $\sim 0.3$~mag.  Our final $z \sim 2.1$ sample then 
consists of 37 LAEs brighter than the Gu10 90\% completeness limit of 
$m_{AB} = 24.8$, and 73 sources belonging to a statistically complete sample 
of objects with monochromatic, continuum-subtracted fluxes greater than
$2 \times 10^{-17}$~ergs~cm$^{-2}$~s$^{-1}$.

\begin{figure}[t]
\figurenum{7}
\plotone{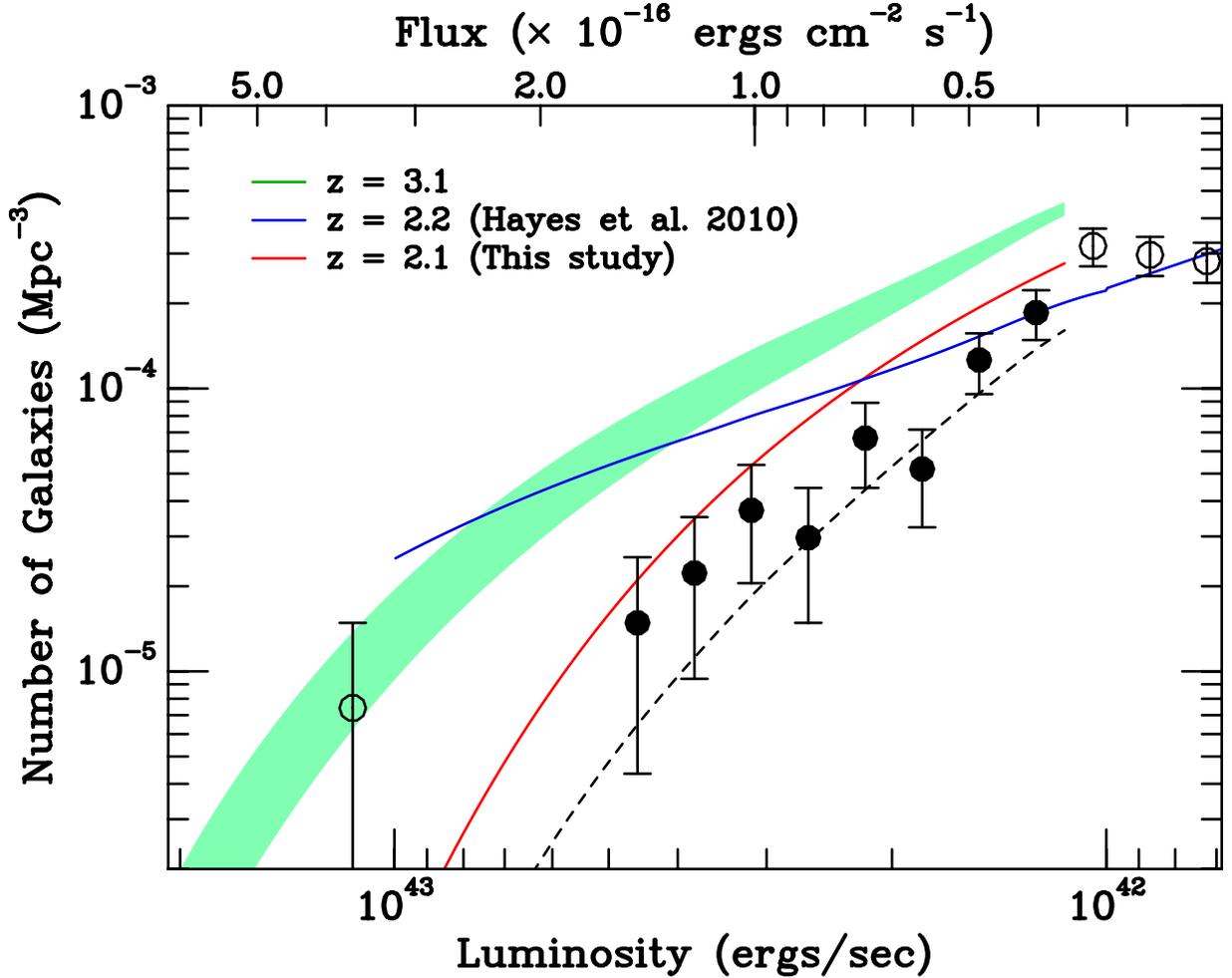}
\caption[lumfun_z2]{The differential luminosity function of $z=2.1$ LAEs with 
rest-frame equivalent widths greater than 20~\AA, inferred from the 
observations of \citet{guaita+10}.   The data have been binned into 0.2~mag 
intervals; points beyond the completeness limit, or where photometric errors 
may cause an excess amount of contamination are shown as open circles.   The 
open circle at the bright end represents the one extremely luminous object 
that has been classified as a $z=2$ QSO by \citet{combo17}.  The solid red 
line is the best-fit \citet{schechter} function, while the dashed line shows 
the shape and normalization of this function after correcting for the effects 
of photometric error and censoring by our filter's non-square transmission 
curve.  The shaded area represents the range of $z \sim 3.1$ luminosity 
functions, and the blue line is the Schechter fit to $z \sim 2.2$ LAEs found by
\citet{hayes}.  The data suggest that $L^*$ has faded by $\sim 0.4$~mag 
between $z \sim 3.1$ and $z \sim 2.1$.}
\label{lumfun_z2}
\end{figure}

In Figure~\ref{lumfun_z2}, we fit the Gu10 LAE monochromatic fluxes (with
the ``large'' objects excluded) to a \citet{schechter} function in exactly 
the same manner as for the $z \sim 3.1$ samples, setting $\alpha = -1.65$, 
and using convolution kernels appropriate for the Gu10 photometric errors and 
filter transmission function.  The first feature to note is the offset between 
the dashed line, which displays the best fit to the data, and the solid black 
line, which represents the implied luminosity function.  This difference, which
is almost 3 times larger than in Figure~\ref{lumfun_z3}, is due entirely to 
the properties of the narrow-band filter employed in the survey.   Like the
Gr07 observations, the Gu10 study was performed through a 
filter whose transmission curve is more Gaussian-shaped than top-hat.
Consequently, LAEs with redshifts that place their emission-line off the
center of the filter have their monochromatic fluxes (and equivalent
widths) systematically underestimated.  The result is an effective 
survey volume that is less than that implied by FWHM of the filter.  
(See Gr07 for a full description of the effect.)  

\begin{figure}[t]
\figurenum{8}
\plotone{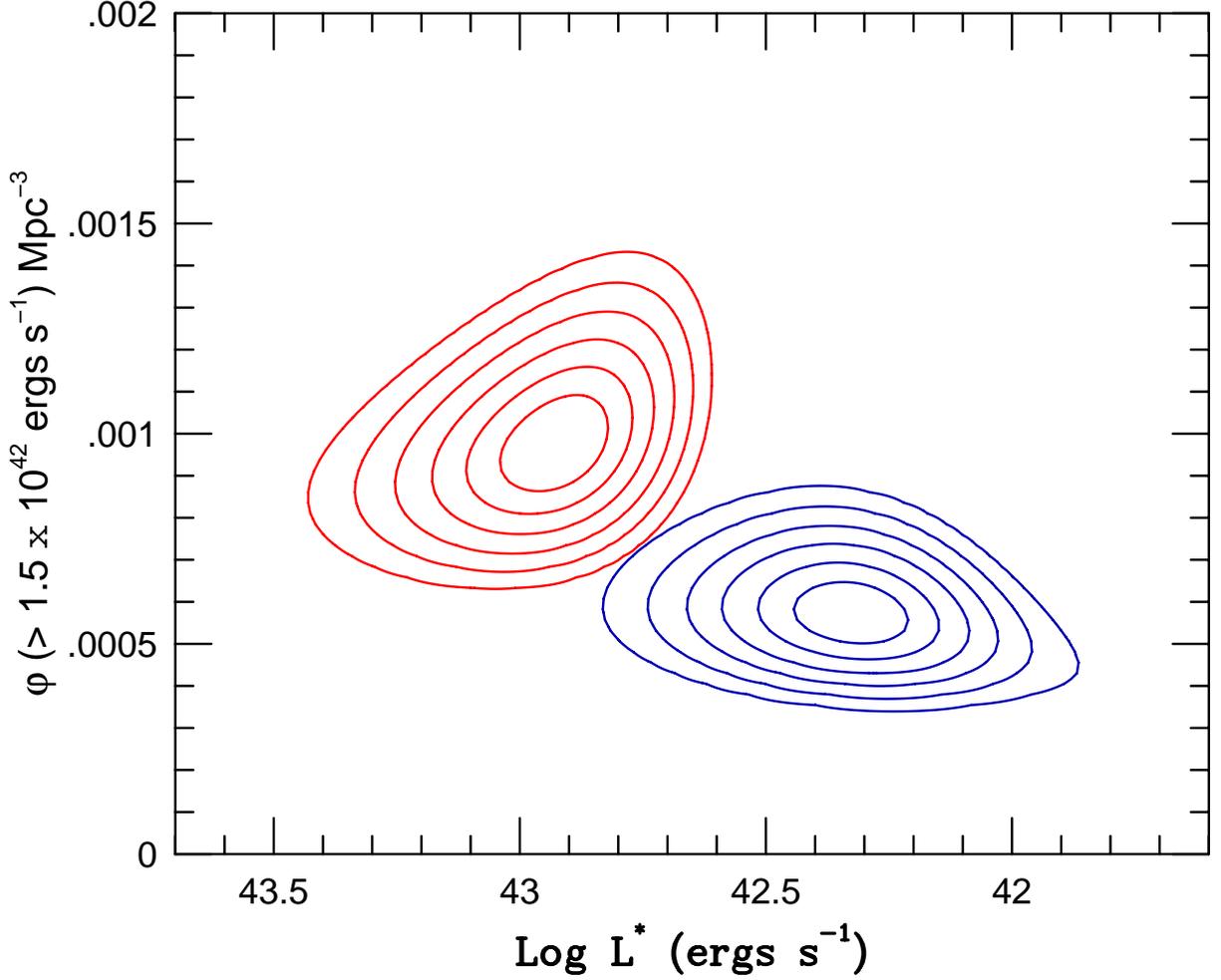}
\caption[2contour]{A comparison of the likelihood confidence contours of 
\citet{schechter} function fits to the observed distribution of Ly$\alpha$ 
luminosities.  The red contours, drawn at $0.5 \, \sigma$ intervals, show the 
results for our $z=3.1$ LAE sample; the blue curves display the same 
information for the Gu10 $z=2.1$ dataset.  The abscissa represents the 
continuum-subtracted value for $\log L^*$; the ordinate displays the integral 
of the \citet{schechter} function down to a luminosity of 
$1.5 \times 10^{42}$~ergs~s$^{-1}$.  Between the two redshifts,
the density of LAEs has decreased by more than 50\%.}
\label{2contour}
\end{figure}

Even after we correct for this offset, it is apparent that there are 
fewer bright LAEs at $z \sim 2.1$ than predicted by the $z \sim 3.1$ 
luminosity functions.  Our $z \sim 2.1$ value of $\log L^* = 42.33 \pm 0.12$, 
is $\sim 0.4$~mag fainter than that measured at the higher redshift, and there 
are $\sim 50\%$ fewer LAEs with emission-line luminosities brighter
than $1.5 \times 10^{42}$~ergs~s$^{-1}$.  Even if the ``large'' LAE 
candidates are included in the fit, the result does not change: the inclusion
of these objects brightens $L^*$ by only 0.01~mag, and increases the 
normalization of the function by only 4\%.  The dramatic change in the 
luminosity function between $z=3.1$ and $z=2.1$ is also illustrated via the 
likelihood contours of Figure~\ref{2contour}, and is similar,
though a bit larger, than the fall-off seen by \citet{blanc} in their blind 
integral-field spectroscopic survey of the COSMOS, GOODS-N, MUNICS-S2, and 
XMM-LSS fields.  We note, however, that though the \citet{blanc} observations 
with the McDonald 2.7-m telescope dwarf all other LAE surveys in terms of 
survey volume $\sim 10^6$~Mpc$^3$), their measurement of LAE evolution is 
limited by the relatively small number of objects detected ($\sim 100$~LAEs 
distributed roughly evenly between $1.9 < z < 3.8$), and a very 
bright limiting luminosity ($\sim 10^{43}$~ergs~s$^{-1}$).  Similarly, our
results are consistent with the $z=2.25$ narrow-band photometry of 
\citet{nilsson09a}, though again, the size of the telescope (2.2~m) and 
width of the interference filter (130~\AA\ FWHM) restricted their study to 
relatively luminous objects.  Still, this survey did detect $\sim 190$~LAE 
candidates with little expected contamination, and the global properties of 
the sample should be robust.  The only apparent contradiction with our data
involes the $z=2.2$ analysis of \citet{hayes}, who used a combination of 
Ly$\alpha$, H$\alpha$, and deep broadband imaging (the latter for 
photo-$z$ estimates) to derive a much slower rate of evolution at the bright 
end of the LAE luminosity function.  However, this result is based on only 
21 LAEs, none brighter than $2 \times 10^{42}$~ergs~s$^{-1}$.
Thus, it is still possible that all the results are consistent.

%It must be pointed out, however, that our most-likely Schechter function
%is not a particularly good fit to the observed $z \sim 2.1$ LAE number
%counts.  Although neither a K-S or $\chi^2$ statistic can rule out consistency
%at more than than $\sim 90\%$ confidence level, neither do they show
%good agreement.  One explanation for this discrepancy is possible contamination
%of the $z \sim 2.1$ LAE sample by foreground or spurious sources.  Gu10
%estimate the fraction of interlopers in their sample to be 
%$\sim 7\% \pm 7\%$.  If contamination is a problem, then we might expect the
%effect to be worse at faint magnitudes, and, perhaps not so coincidentally, 
%it is the last bin of the luminosity function that contributes 
%most to the $\chi^2$ deviation.  This suggests that the decrease in
%the LAE number density towards lower redshifts might be slightly greater
%than that measured.  

\begin{figure}[t]
\figurenum{9}
\plotone{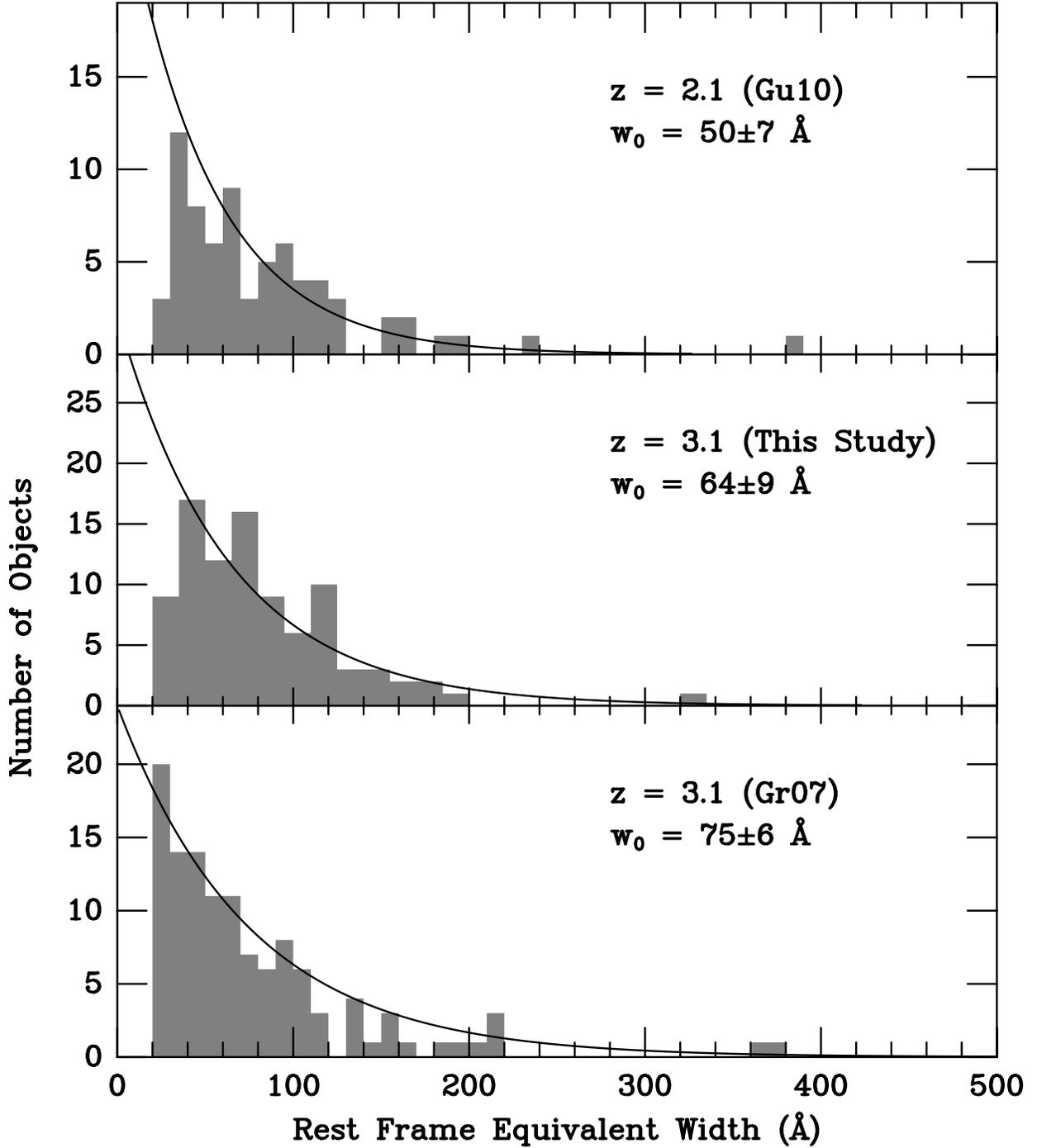}
\caption[ew_compare]{Equivalent width histograms for our new sample of 
$z \sim 3.1$ LAEs, compared to that measured by Gr07 (for $z \sim 3.1$) and 
Gu10 (for $z \sim 2.1$).  The curves are the best-fitting exponentials,
corrected for the effects of photometric error and censoring by the filters'
non-square transmission curves.  There are significantly fewer high 
equivalent width objects at $z \sim 2.1$ than at $z \sim 3.1$.}
\label{ew_compare}
\end{figure}

More evidence of LAE evolution comes from the distribution of Ly$\alpha$
equivalent widths.  Figure~\ref{ew_compare} displays a histogram of 
rest-frame equivalent widths for the present $z=3.1$ sample, the
Gr07 $z=3.1$ sample, and the Gu10 $z=2.1$ dataset.
The two $z=3.1$ distributions are statistically identical: K-S
and Anderson-Darling tests can discern no difference between the datasets,
and an exponential fit to the distributions, using 
the same maximum-likelihood procedures as Gr07, yields a rest-frame 
scale-length of $w_0 = 64^{+10}_{-7}$~\AA\ for the LAEs of our new survey.
This is consistent with the value of $w_0 = 76^{+11}_{-8}$~\AA\ found
by Gr07, and, when the two datasets are combined, the resulting
equivalent width scale length for $z=3.1$~LAEs is $70^{+7}_{-5}$~\AA.
This contrasts sharply with the best-fit value of $w_0 = 50^{+9}_{-6}$~\AA\ 
for the $z \sim 2.1$~LAE sample.  Non-parametric statistics, such
as those associated with the K-S and Anderson-Darling tests, prove that the
samples differ with greater than 99\% confidence, confirming that some
amount of evolution has occurred in the Gyr between $z=3.1$ and $z=2.1$.

\section{Ly$\alpha$ and UV Star Formation Rates}
The strength of the Ly$\alpha$ emission line and brightness of the
UV continuum are both functions of star formation.   Specifically,
the assumptions of an optically thick interstellar medium, Case~B
recombination, and a \citet{salpeter} universal initial mass function
lead to a relationship between Ly$\alpha$ luminosity (in ergs~s$^{-1}$)
and star-formation rate, 
\begin{equation}
{\rm SFR(Ly}\alpha) = 9.1 \times 10^{-43} \, L({\rm Ly}\alpha) \
M_{\odot}~{\rm yr}^{-1}
\label{eq_sfr_la}
\end{equation}
\citep{kennicutt, brocklehurst}.  Similarly, population synthesis 
models for systems of stars undergoing a constant rate of star formation
predict 
\begin{equation}
{\rm SFR(UV)} = 1.4 \times 10^{-28} \, L_{\nu} \ M_{\odot}~{\rm yr}^{-1}
\label{eq_sfr_uv}
\end{equation}
where $L_{\nu}$ (in ergs~s$^{-1}$~Hz$^{-1}$) represents the population's 
mid-UV (1500~\AA\ to 2800~\AA) continuum luminosity \citep{kennicutt}.
Thus, in the absence of dust, galaxies which are 
optically thick to Lyman radiation should have Ly$\alpha$ luminosities that 
are well-correlated with their continuum flux.

\begin{figure}[t]
\figurenum{10}
\plotone{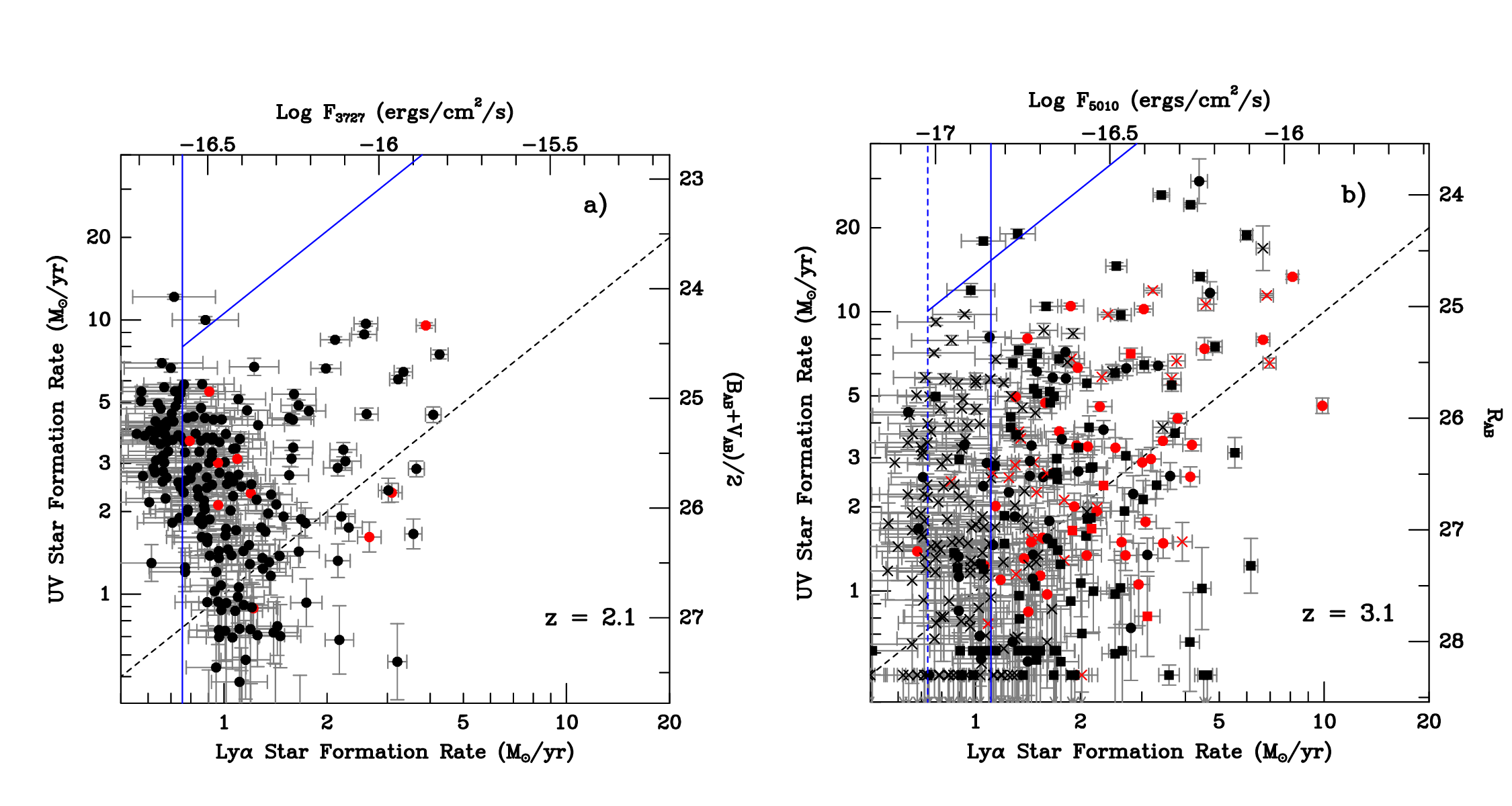}
%\hskip1.2in
%\includegraphics[angle=90, clip=true]{f10.eps}
\caption[sfr]{A comparison of the star formation rates derived from Ly$\alpha$ 
emission (under Case B recombination) and the UV continuum at 1570~\AA.  The
left-hand panel shows $z \sim 2.1$ LAEs from Gu10, while the right-hand panel
displays $z \sim 3.1$ objects, with crosses representing LAEs from
Gr07, squares denoting the sources found in the current study, and circles
showing candidates present in both surveys.  The red points show
LAEs that have been spectroscopically confirmed.  The diagonal
dashed line shows where the two measurements are equal.  The flux limits
are shown via the vertical blue line (in the case of the 
$z=3.1$ data, solid for this survey, dashed for Gr07); the approximate 
location of our equivalent width threshold is shown via the diagonal
blue line.  Note that, although the two indicators are correlated, 
there is a substantial amount of scatter in the relation.}
\label{sfr}
\end{figure}

Figure~\ref{sfr} compares the two quantities.  From the figure, it is easy
to see that at both redshifts, there is a wide range of star-formation
rate ratios, and that the scatter in the diagram is much larger than 
that produced by simple photometric errors.  Objects above the
one-to-one line are easily explained by the loss of Ly$\alpha$ photons,
most likely to encounters with dust grains \citep[\eg][]{charlot93,
hayes, atek09}.  In contrast, objects below the line may be systems for which
the ``continuous star formation approximation'' in population synthesis
models is inappropriate (\ie\ objects with star-formation timescales less 
than $\sim 10^8$~yr) or ones where the dust is preferentially embedded 
within large neutral gas clouds with steep density profiles 
\citep[\eg][]{neufeld, hansen, finkelstein09}.

\begin{figure}[t]
\figurenum{11}
\plotone{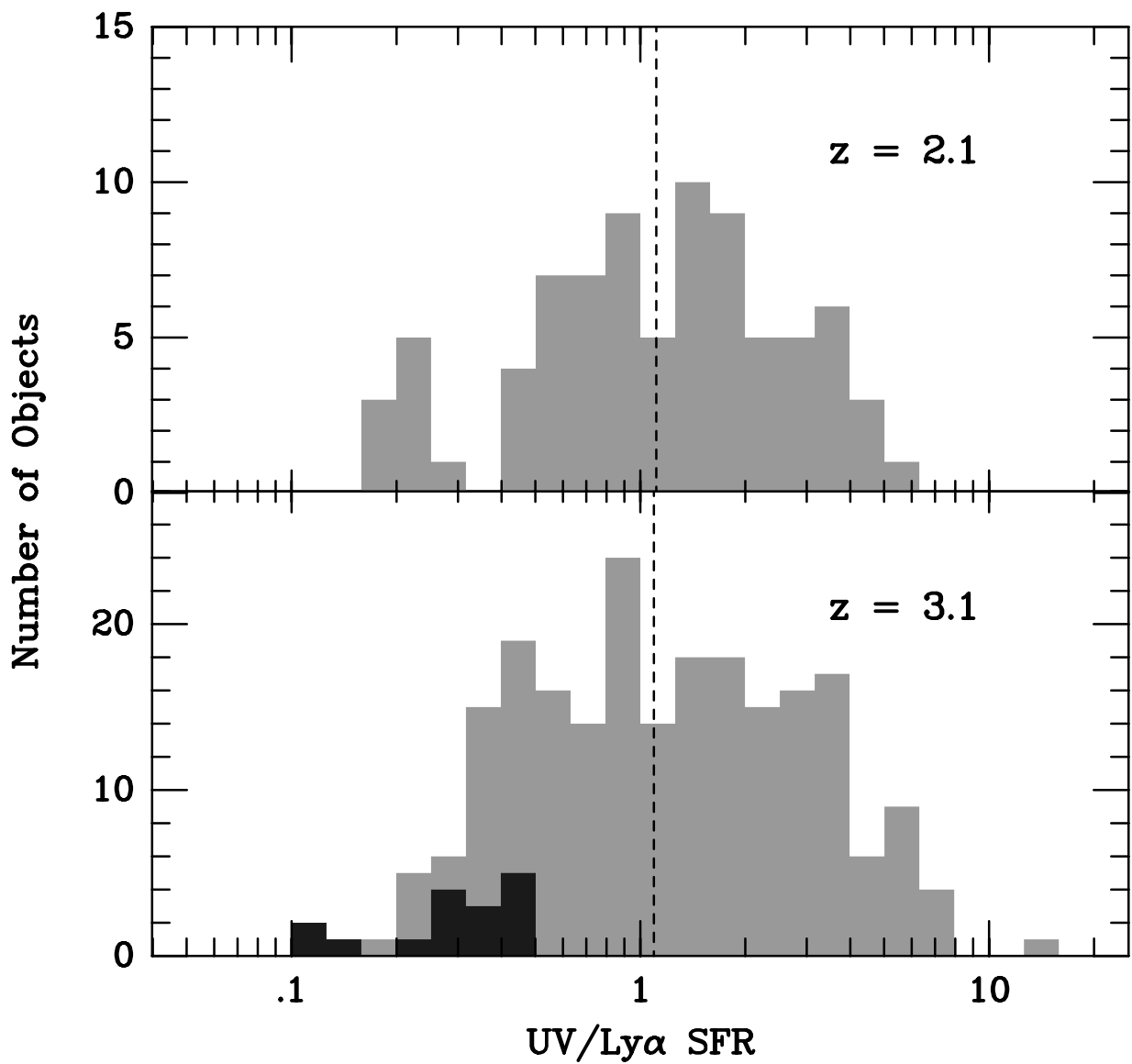}
\caption[sfr_hist]{The ratio of the star-formation rate derived from the UV 
continuum to that derived from Ly$\alpha$ (under Case~B recomination) for all 
LAEs with line luminosities brighter than $\log L = 42.08$.  The dark bins 
represent objects for which we only have upper limits; the dashed lines show 
the median values of the distributions: 1.09 for $z = 3.1$ LAEs and 
1.11 for the $z=2.1$ objects.  An Anderson-Darling test cannot detect
any difference between the two distributions.}
\label{sfr_hist}
\end{figure}

Interestingly, the distribution of star-formation rate ratios displayed 
in Figure~\ref{sfr} shows no evidence for evolution between $z \sim 2.1$
and $z \sim 3.1$.  Our new $z=3.1$ survey is complete to a 
continuum-subtracted monochromatic luminosity limit of 
$1.2 \times 10^{42}$~ergs~s$^{-1}$.  For LAEs brighter than this, the median
ratio of the UV star-formation rate to the Ly$\alpha$ star-formation rate
is $1.09^{+0.19}_{-0.08}$, where the uncertainty is determined via a bootstrap
analysis.  For $z\sim 2.1$ systems, this ratio is $1.11^{+0.20}_{-0.15}$.
Moreover, as Figure~\ref{sfr_hist} illustrates, there is no difference
in the shape of the underlying distribution:  while a number of physical
mechanisms might cause a systematic shift in the ratio of the two 
star formation rate indicators, an Anderson-Darling test can find no 
evidence for any evolution, even at the $2 \, \sigma$ level.  

Nevertheless, the change in the equivalent width distribution, does point 
to a decrease in the efficiency of Ly$\alpha$ emission.  As described 
above, Ly$\alpha$ line flux and mid-UV continuum flux are both driven 
by a galaxy's star-formation rate.  Moreover, in the wavelength range 
$1250~{\rm\AA} < \lambda < 2600~{\rm\AA}$, a star-forming galaxy's
spectral energy distribution (SED) is well-fit by a simple power-law, 
$F_{\lambda} \propto \lambda^{\beta}$, with slope $\beta \sim -2.3$
\citep{calzetti+94, meurer+99}.  Since $\beta$ is rather insensitive 
to a starburst's duration, metallicity, or initial mass function 
\citep{leitherer99, bouwens+09}, any systematic shift in a population's
Ly$\alpha$ to 1250~\AA\ continuum flux ratio, \ie\ in its 
Ly$\alpha$ equivalent width, must be due to some other parameter.
Geometry is one such variable, since an increase in the opening angle
for Ly$\alpha$ emission would cause a typical line-of-sight to receive
fewer photons.  However, purely geometrical effects cannot explain
the fading of the LAE luminosity function, nor a decline in the
star formation rate density of the LAE population (see next section).
The most likely explanation for these changes is dust, which can 
decrease $\beta$, and also reduce the efficiency of line emission by 
destroying Ly$\alpha$ photons.  This interpretation is supported by
the SED analysis of \citet{guaita+11}, who found that $z=2.1$ LAEs have
slightly greater reddening than their $z=3.1$ counterparts 
\citep{gawiser07}.

\section{The Star Formation Rate Density of LAEs}
\begin{figure}[t]
\figurenum{12}
\plotone{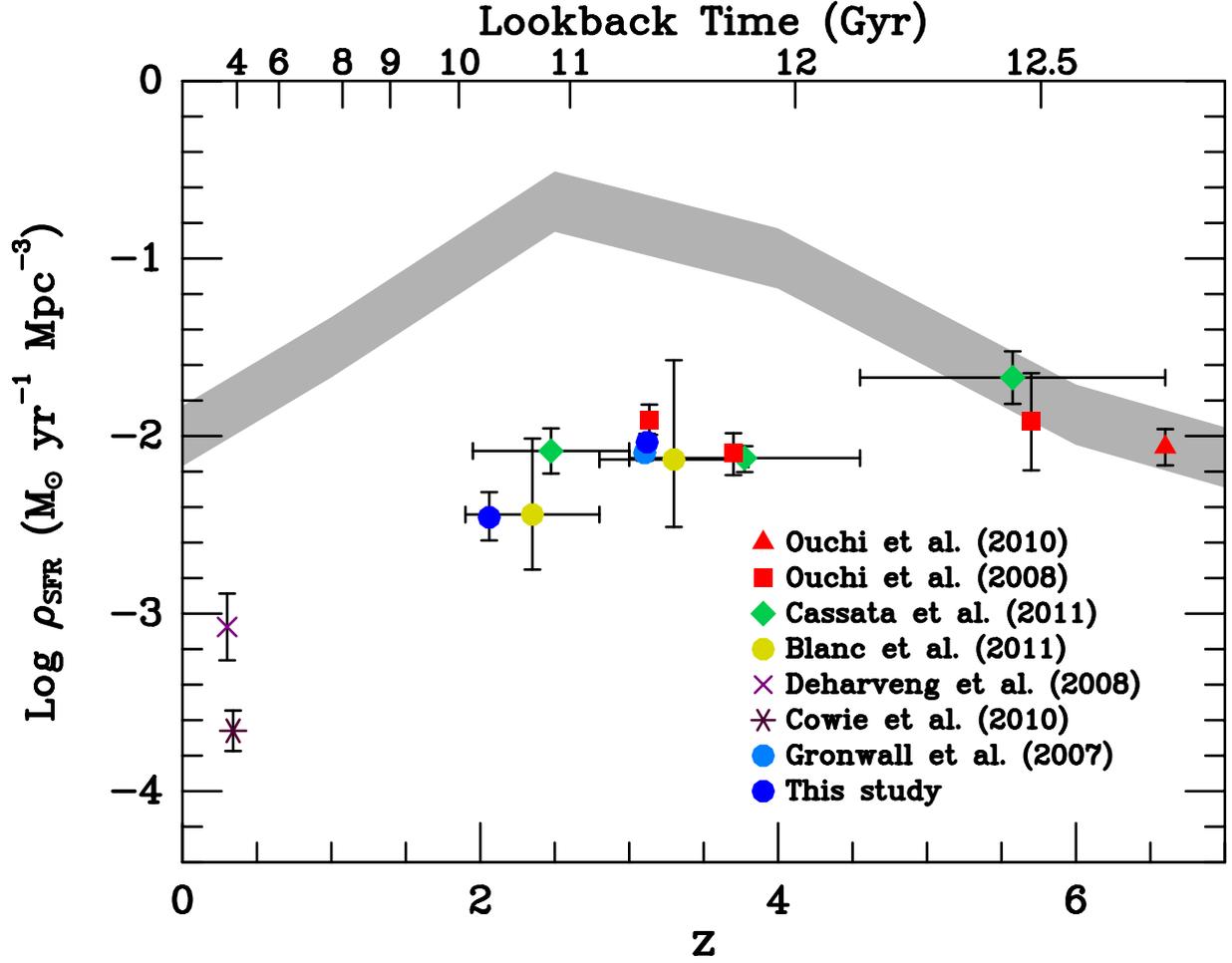}
\caption[sfr_density]{Measurements of the observed LAE star-formation rate 
density as a function of redshift.  Each point is derived from by integrating 
the Schechter function down to zero luminosity, and for consistency, each
assumes a faint-end slope of $\alpha = -1.65$.  The shaded area displays the 
results from the rest-frame UV \citep{bouwens+10}.  At $z \gtrsim 6$, the LAE 
and UV-based measurements of the star-formation rate density are equal, but by
$z \sim 4$, this ratio has transitioned to a value of $\lesssim 0.1$.}
\label{sfr_density}
\end{figure}

Narrow-band imaging \citep[][Gr07, Gu10]{ouchi+08, ouchi+10},
along with slitless \citep{deharveng}, long-slit \citep{cassata},
and integral-field \citep{blanc} spectroscopy now yield a 
coherent picture of the evolution of the LAE population and their
star-formation rate density.  To illustrate this, we computed the
total observed Ly$\alpha$ luminosities of the $z=2.1$ and $z=3.1$
LAE populations by integrating their best-fit \citet{schechter} 
functions of Table~\ref{sch_lf} down to zero.  We then used 
equation~(\ref{eq_sfr_la}) to convert these luminosities into 
star-formation rate densities, and compared our results to those of
other LAE measurements and from estimates made using the rest-frame
UV continuum.  This comparison is shown in Figure~\ref{sfr_density}.
As the figure indicates, at $z \gtrsim 5$, the star-formation rate
densities one obtains from Ly$\alpha$ are identical to those derived 
from the rest-frame UV; at these redshifts, the assumption of an
optically thick Case~B interstellar medium, little extinction, and
a Ly$\alpha$ escape fraction near unity appears reasonable.  
However, at a lookback time of $\sim 12$~Gyr, there is 
a rapid transition, as at all subsequent epochs the ratio of LAE to
rest-frame UV star-formation rate density is $\lesssim 0.1$. 
The decline in the Ly$\alpha$-based measurements between $z \sim 3.1$
and $z \sim 2.1$ (and, indeed, to $z \sim 0$) roughly follows that 
for the universe as a whole.

Figure~\ref{sfr} shows that, in a differential sense, dust has very little
effect on the $2 \lesssim z \lesssim 4$ LAE population:  on average, the 
star-formation rates inferred from Ly$\alpha$ are similar to those derived 
from the rest-frame UV\null.  The dispersion in the diagram is large, 
suggesting the presence of some reddening (and perhaps implying
that many of the LAEs have star-formation timescales less than
$\sim 10^8$~yr), but overall, the Ly$\alpha$ escape fraction from
these objects appears to be close to unity.  On the other hand,
Figure~\ref{sfr_density} proves that globally, Ly$\alpha$-based
measurements of star formation yield values that are more than an
order of magnitude less than those found from the rest-frame UV\null.
Clearly, the bulk of star formation is occuring outside of the LAE
population.  As confirmed by their colors and nebular emission,
LAEs represent the extreme low-extinction, low-metallicity tail of the 
star-forming galaxy population \citep{gawiser07, guaita+11, 
finkelstein+11}.

\section{Conclusion}
Our 5010~\AA\ narrow-band survey of the Extended Chandra Deep Field South 
has produced 199 high equivalent width (EW $> 80$~\AA) objects, of which
141 are members of a statistically complete sample with monochromatic
fluxes brighter than $2.4 \times 10^{-17}$~ergs~cm$^{-2}$~s$^{-1}$.  
X-ray and UV observations with the Chandra, GALEX, and Swift satellites 
demonstrate that most of our brightest candidates are AGN or foreground [O~II]
emitters, but when we exclude these sources, we are still left with a
complete sample of 130 candidates that are likely to be Ly$\alpha$ emitting
galaxies at $z \sim 3.12$.  This sample overlaps that defined by
Gr07 using a 4990 \AA\ narrow-band filter, but includes an additional
68 candidates.  This brings the total number of likely $z \sim 3.1$ LAEs
known in the ECDF-S to 360, with 223 being members of statistically complete
samples.  The sample now extends in redshift space from $z = 3.06$ to
$z \sim 3.14$ and includes 72 objects with spectroscopic confirmations.

Using these data, we have explored the luminosity function and
equivalent width distribution of $z \sim 3.1$ LAEs.  Both functions
are in excellent agreement with previous determinations:  our
value for the Schechter cutoff ($\log L^* = 42.76 \pm 0.10$)
and equivalent width e-folding scale ($w_0 = 64^{+10}_{-7}$~\AA) are
statistically identical to values in the literature 
\citep[Gr07,][]{ouchi+08} and the normalization of the functions 
are also consistent to $\sim 25\%$.  We show that an apparent
apparent deficit of continuum-bright, high equivalent-width objects is 
most likely an artifact caused by correlated errors in our
heteroskedastic dataset.  When the independent parameters of
line-luminosity and continuum luminosity are compared, the 
apparent correlation disappears.  

We have also examined the evolution of the LAE luminosity function
towards lower redshifts by comparing our sample of $z \sim 3.1$ 
Ly$\alpha$ emitters to the $z \sim 2.1$ LAEs found by Gu10\null.  We show
that in the $\sim 1$~Gyr between the two epochs, the number of LAEs
with line luminosities brighter than $1.5 \times 10^{42}$~ergs~s$^{-1}$ 
declined by $\sim 50\%$, and $\log L^*$ faded by $\sim 0.4$~mag.
Thus, it appears that the large decrease in the frequency and strength
of galactic Ly$\alpha$ emission observed by \citet{deharveng} and
\citet{cowie} at $z \sim 0.3$ has begun by $z \sim 2$.  Similarly, 
there are significantly fewer $z \sim 2.1$ LAEs with extremely high
equivalent widths:  between $z \sim 3.1$ and $z \sim 2.1$, the
scale length of the LAE equivalent width distribution decreased from
$70^{+7}_{-5}$~\AA\ to $50^{+10}_{-7}$~\AA.  This decrease suggests
that $z \sim 2.1$ LAEs have slightly more dust than their $z \sim 3.1$
counterparts.  This hypothesis is supported by the results of
spectral energy distribution fitting \citep{gawiser07, guaita+11}, but 
is not confirmed by an analysis of the objects' UV to Ly$\alpha$-based
star formation rate ratios.  

Finally, by integrating our LAE luminosity functions and then combining
our measurements with other data from the literature, we have estimated how 
the contribution of LAEs to the star-formation rate density of the universe
changes with redshift.  At $z \gtrsim 5$, LAEs are responsible for
most of the star-formation rate density, but in the redshift range 
$2 \lesssim z \lesssim 3$, this fraction has decreased to $\sim 0.1$.
As pointed out by \citet{blanc}, this suggests that globally, there has been 
strong evolution in the Ly$\alpha$ escape fraction.  This is consistent with 
other results based on LAE photometry and spectroscopy:  at the epochs 
considered here, Ly$\alpha$ emitters represent the extreme low-metallicity, 
low-extinction, faint-end tail of the star-forming galaxy population 
\citep{gawiser07, guaita+11, finkelstein+11}.

\acknowledgements
We would like to thank Erik Hoversten for his assistance with the
Swift-UVOT catalog, Xuesong Wang for her help with EIS J033330.60-274819.3,
and Guillermo Blanc for providing 
the results of the HETDEX pilot survey prior to publication.  
This work was supported by NSF grants 
AST 06-07416, AST 08-07570, and AST 08-07885, and DOE grants 
DE-GF02-08ER41560 and DE-FG02-08ER41561.  

{\it Facilities:} \facility{Blanco (Mosaic)}

\clearpage

%\figcaption[venn]
%{A schematic showing the relationship between Lyman break galaxies
%and Ly$\alpha$ emitters.  The dotted diagonal line represents
%galaxies in which the Ly$\alpha$ line has zero equivalent width.
%Since LBG surveys are conducted with broadband filters, they can only 
%detect objects brighter than some (rest-frame UV) magnitude limit.  
%Conversely, LAEs are identified via the contrast of their 
%emission-line over the continuum, so these objects are largely defined
%via a limiting equivalent width.  Note that faint galaxies with 
%weak emission cannot be seen with either technique.  The dark
%triangular region represents systems with unrealistically broad
%Ly$\alpha$ absorption.
%\label{venn}
%}

\clearpage

\begin{deluxetable}{lccl}
\tablewidth{0pt}
\tablecaption{Log of Narrow-band Observations}
\tablehead{&\colhead{Exposure} &\colhead{Seeing} \\
\colhead{UT Date} &\colhead{(min)} &\colhead{($\arcsec$)} &\colhead{Notes}}
\startdata
15 Oct 2006   &$9 \times 20$   &$1\farcs 1$  &Through clouds \\
17 Oct 2006   &$8 \times 20$   &$1\farcs 1$  &Photometric \\
19 Oct 2006   &$11 \times 20$  &$1\farcs 1$  &Photometric \\
21 Oct 2006   &$12 \times 20$  &$1\farcs 1$  &Photometric \\
20 Nov 2006   &$3 \times 20$   &$1\farcs 1$  &Photometric \\
22 Nov 2006   &$4 \times 20$   &$1\farcs 2$  &Light cirrus \\
\enddata
\label{obslog}
\end{deluxetable}
\clearpage

\begin{deluxetable}{lcccccl}
\tablewidth{0pt}\tabletypesize{\scriptsize}
\tablecaption{Candidate Ly$\alpha$ Emitters: The Statistically Complete Sample}
\tablehead{
\colhead{ID}
&\colhead{$\alpha$(2000)}
&\colhead{$\delta$(2000)}
&\colhead{Log $F_{5010}$}
&\colhead{Observed EW (\AA)}
&\colhead{$z$}
&\colhead{Comments\tablenotemark{a}}  }
\startdata
  1 &3:32:17.57 &$-$27:49:40.9 &$-15.227$ &  98 &0.337 &GALEX, UVOT source; Resolved\\
  2 &3:32:04.06 &$-$27:37:25.4 &$-15.523$ & 124 &0.977 &x-ray, GALEX source\\
  3 &3:31:46.18 &$-$27:57:08.6 &$-15.523$ & 129 &\dots &x-ray, GALEX, UVOT source\\ 
  4 &3:33:07.61 &$-$27:51:26.9 &$-15.707$ & 153 &1.609 &x-ray, UVOT source; \\
  5 &3:31:40.14 &$-$28:03:07.5 &$-15.723$ & 249 &\dots &G-009           \\
  6 &3:31:34.73 &$-$27:56:21.8 &$-15.743$ & 377 &3.131 &G-019           \\
  7 &3:33:12.72 &$-$27:42:46.9 &$-15.767$ & 389 &\dots &G-002           \\
  8 &3:31:51.04 &$-$27:57:15.7 &$-15.771$ & 195 &\dots &x-ray, GALEX source \\
  9 &3:33:04.86 &$-$27:45:54.9 &$-15.811$ &  97 &\dots &                 \\
 10 &3:32:18.80 &$-$27:42:48.2 &$-15.851$ & 230 &3.116 &G-004            \\
\enddata
\tablenotetext{a}{G-XXX denotes identification number in \citet{g+07}}
\label{brightLAEs}
\end{deluxetable}
\clearpage

\begin{deluxetable}{lcccccl}
\tablewidth{0pt}\tabletypesize{\scriptsize}
\tablecaption{Candidate Ly$\alpha$ Emitters: Objects Below the
Completeness Limit}
\tablehead{
\colhead{ID}
&\colhead{$\alpha$(2000)}
&\colhead{$\delta$(2000)}
&\colhead{Log $F_{5010}$}
&\colhead{Observed EW (\AA)}
&\colhead{$z$}
&\colhead{Comments\tablenotemark{a}}  }
\startdata
 142 &3:32:06.86 &$-$27:53:33.4 &$-16.619$ & 611 &\dots  &                 \\
 143 &3:32:58.57 &$-$27:41:31.0 &$-16.619$ & 307 &\dots  &G-221            \\
 144 &3:31:52.83 &$-$27:45:18.6 &$-16.623$ & 186 &3.106  &G-050            \\
 145 &3:33:12.89 &$-$28:03:08.2 &$-16.627$ & 739 &\dots  &                 \\
 146 &3:33:18.56 &$-$27:33:49.5 &$-16.635$ & 926 &\dots  &                 \\
 147 &3:33:21.69 &$-$27:45:06.9 &$-16.635$ & 163 &\dots  &                 \\
 148 &3:33:02.99 &$-$28:02:34.9 &$-16.635$ & 136 &\dots  &                 \\
 149 &3:32:57.35 &$-$27:51:42.1 &$-16.635$ & 331 &\dots  &G-102            \\
 150 &3:33:38.74 &$-$27:47:02.5 &$-16.643$ & 460 &\dots  &G-142            \\
 151 &3:32:14.57 &$-$27:45:52.4 &$-16.647$ & 314 &3.118  &G-090            \\
\enddata
\tablenotetext{a}{G-XXX denotes identification number in \citet{g+07}}
\label{faintLAEs}
\end{deluxetable}
\clearpage

\begin{deluxetable}{cc|cc}
\tablewidth{0pt}
\tablecaption{Photometric Uncertainties}
\tablehead{
\colhead{Log $F_{5010}$} &\colhead{$\sigma$ (mag)}
&\colhead{Log $F_{5010}$} &\colhead{$\sigma$ (mag)} }
\startdata
$-15.50$    &0.012    &$-16.30$   &0.079  \\
$-15.60$    &0.016    &$-16.40$   &0.095  \\
$-15.70$    &0.022    &$-16.50$   &0.115  \\
$-15.80$    &0.029    &$-16.60$   &0.134  \\
$-15.90$    &0.036    &$-16.70$   &0.167  \\
$-16.00$    &0.044    &$-16.80$   &0.198  \\
$-16.10$    &0.052    &$-16.90$   &0.235  \\
$-16.20$    &0.066    &$-17.00$   &0.288  \\
\enddata
\label{errors}
\end{deluxetable}

\begin{deluxetable}{lcccc}
\tablewidth{0pt}
\tablecaption{Best-Fit Schechter Function Parameters}
\tablehead{
&\colhead{$\log L^*$}  &&\colhead{$\log \phi^*$} 
&\colhead{$\log \rho_{{\rm Ly}\alpha}^{\rm tot}$ } \\
\colhead{Sample} &(ergs~s$^{-1}$) &\colhead{$\alpha$} 
&\colhead{(Mpc$^{-3}$)}  &(ergs~s$^{-1}$~Mpc$^{-3}$) }
\startdata
$z = 3.095$ (Gr07) &$42.70 \pm 0.10$  &$-1.65$  &$-3.14 \pm 0.04$ 
&$39.97 \pm 0.11$ \\
$z = 3.113$        &$42.76 \pm 0.10$  &$-1.65$  &$-3.17 \pm 0.05$
&$39.99 \pm 0.11$ \\
$z = 3.1$ \citep{ouchi+08}\tablenotemark{a} 
&$42.82 \pm 0.06$ &$-1.65$ &$-3.15 \pm 0.11$
&$40.08 \pm 0.13$ \\
$z = 2.063$  (Gu10) &$42.33 \pm 0.12$  &$-1.65$ &$-2.86 \pm 0.05$
&$39.59 \pm 0.13$ \\
\enddata
\tablenotetext{a}{Values are interpolated to $\alpha = -1.65$, and appropriate
for a 64~\AA\ rest-frame equivalent width limit.  Following their suggestion, 
these numbers should be increased by $\sim 10\%$ to approximate our 20~\AA\
limit.}
\label{sch_lf}
\end{deluxetable}

\clearpage


\clearpage
\begin{thebibliography}{}

\bibitem[Albrecht \etal(2006)]{detf} Albrecht, A., Bernstein, G., Cahn, R., 
Freedman, W.L., Hewitt, J., Hu, W., Huth, J., Kamionkowski, M., Kolb, E.W., 
Knox, L., Mather, J.C., Staggs, S., \& Suntzeff, N.B. 2006, 
arXiv:astro-ph/0609591 

\bibitem[Ando \etal(2006)]{ando+06} Ando, M., Ohta, K., Iwata, I.,
Akiyama, M., Aoki, K., \& Tamura, N. 2006, \apjl, 645, L9 

\bibitem[Ando \etal(2007)]{ando+07} Ando, M., Ohta, K., Iwata, I.,
Akiyama, M., Aoki, K., \& Tamura, N. 2007, \pasj, 59, 717 

\bibitem[Atek \etal(2009)]{atek09} Atek, H., Kunth, D., Schaerer, D., 
Hayes, M., Deharveng, J.M., \"Ostlin, G., \& Mas-Hesse, J.M. 2009, 
\aap, 506, L1

\bibitem[Balestra \etal(2010)]{balestra} Balestra, I.,  Mainieri, V., 
Popesso, P., Dickinson, M., Nonino, M., Rosati, P., Teimoorinia, H., 
Vanzella, E., Cristiani, S., Cesarsky, C., Fosbury, R.A.E., Kuntschner, H., 
\& Rettura, A.  2010, \aap, 512, A12

\bibitem[Blanc \etal(2011)]{blanc} 
Blanc, G.A., Adams, J., Gebhardt, K., Hill, G.J., Drory, N., Hao, L.,
Bender, R., Byun, J., Ciardullo, R., Finkelstein, S., Fry, A., 
Gawiser, E., Gronwall, C., Hopp, U., Jeong, D., Kelz, A., Kelzenberg, R.,
Komatsu, E., MacQueen, P., Murphy, J., Palunas, P., Roth, M., 
Schneider, D., \& Tufts, J. 2010, \apj, in press

%\bibitem[Bond \etal(2009)]{bond09} Bond, N.A., Gawiser, E., 
%Gronwall, C., Ciardullo, R., Altmann, M., \& Schawinski, K. 2009, 
%\apj, 705, 639 

%\bibitem[Bond \etal(2010)]{bond10} Bond, N.A., Feldmeier, J.J.,
%Matkovi\'c, A., Gronwall, C., Ciardullo, R., \& Gawiser, E. 2010, 
%\apjl, 716, L200 

\bibitem[Bond \etal(2011)]{bond11} Bond, N., Gawiser, E., 
Guaita, L., Padilla, N., Gronwall, C., Ciardullo, R., 
\& Lai, K.  2011, submitted to Ap.~J. (arXiv:1104.2880)

\bibitem[Bouwens \etal(2009)]{bouwens+09} Bouwens, R.J., Illingworth, G.D., 
Franx, M., Chary, R.-R., Meurer, G.R., Conselice, C.J., Ford, H., 
Giavalisco, M., \& van Dokkum, P. 2009, \apj, 705, 936 

\bibitem[Bouwens \etal(2010)]{bouwens+10} Bouwens, R.J., Illingworth, G.D., 
Oesch, P.A., Stiavelli, M., van Dokkum, P., Trenti, M., Magee, D., 
Labb\'e, I., Franx, M., Carollo, C.M., \& Gonzalez, V. 2010, \apjl, 709, L133 

\bibitem[Brandt \etal(2001)]{brandt} Brandt, W.N., Alexander, D.M., 
Hornschemeier, A.E., Garmire, G.P., Schneider, D.P., Barger, A.J., 
Bauer, F.E., Broos, P.S., Cowie, L.L., Townsley, L.K., Burrows, D.N., 
Chartas, G., Feigelson, E.D., Griffiths, R.E., Nousek, J.A., \&
Sargent, W.L.W. 2001, \aj, 122, 2810 

\bibitem[Brocklehurst(1971)]{brocklehurst} Brocklehurst, M. 1971, 
\mnras, 153, 471 

\bibitem[Calzetti \etal(1994)]{calzetti+94} Calzetti, D., Kinney, A.L., 
\& Storchi-Bergmann, T. 1994, \apj, 429, 582 

\bibitem[Cassata \etal(2011)]{cassata} Cassata, P.,  Le F\`evre, O., 
Garilli, B., Maccagni, D., Le Brun, V., Scodeggio, M., Tresse, L., 
Ilbert, O., Zamorani, G., Cucciati, O., Contini, T., Bielby, R., 
Mellier, Y., McCracken, H.J., Pollo, A., Zanichelli, A., Bardelli, S., 
Cappi, A., Pozzetti, L., Vergani, D., \& Zucca, E. 2011, \aap, 525, A143

\bibitem[Charlot \& Fall(1993)]{charlot93} Charlot, S., \& Fall, S.M.
1993, \apj, 415, 580 

\bibitem[Cowie \etal(2010)]{cowie} Cowie, L.L., Barger, A.J.,
\& Hu, E.M. 2010, \apj, 711, 928 

\bibitem[Deharveng \etal(2008)]{deharveng} Deharveng, J.-M., Small, T., 
Barlow, T.A., P\'eroux, C., Milliard, B., Friedman, P.G., Martin, D.C., 
Morrissey, P., Schiminovich, D., Forster, K., Seibert, M., Wyder, T.K., 
Bianchi, L., Donas, J., Heckman, T.M., Lee, Y.-W., Madore, B.F., 
Neff, S.G., Rich, R.M., Szalay, A.S., Welsh, B.Y., \& Yi, S.K. 2008,
\apj, 680, 1072 

\bibitem[Efron \& Petrosian(1992)]{efron} Efron, B., \& Petrosian, V.
1992, \apj, 399, 345 

\bibitem[Feldmeier \etal(2003)]{ipn2} Feldmeier, J.J., Ciardullo, R.,
Jacoby, G.H., \& Durrell, P.R. 2003, \apjs, 145, 65

\bibitem[Ferguson \etal(2004)]{ferguson04} Ferguson, H.C., Dickinson, M., 
Giavalisco, M., Kretchmer, C., Ravindranath, S., Idzi, R., Taylor, E., 
Conselice, C.J., Fall, S.M., Gardner, J.P., Livio, M., Madau, P., 
Moustakas, L.A., Papovich, C.M., Somerville, R.S., Spinrad, H., \&
Stern, D. 2004, \apjl, 600, L107 

\bibitem[Finkelstein \etal(2009)]{finkelstein09} Finkelstein, S.L., 
Malhotra, S., Rhoads, J.E., Hathi, N.P., \& Pirzkal, N. 2009, 
\mnras, 393, 1174 

\bibitem[Finkelstein \etal(2011)]{finkelstein+11} Finkelstein, S.L., 
Hill, G.J., Gebhardt, K., Adams, J., Blanc, G.A., Papovich, C., 
Ciardullo, R., Drory, N., Gawiser, E., Gronwall, C., Schneider, D.P., 
\& Tran, K.-V. 2011, \apj, 729, 140

\bibitem[Gawiser \etal(2006a)]{gawiser06} Gawiser, E., van Dokkum, P.G., 
Gronwall, C., Ciardullo, R., Blanc, G.A., Castander, F.J., Feldmeier, J., 
Francke, H., Franx, M., Haberzettl, L., Herrera, D., Hickey, T., 
Infante, L., Lira, P., Maza, J., Quadri, R., Richardson, A., Schawinski, K., 
Schirmer, M., Taylor, E.N., Treister, E., Urry, C.M., \& Virani, S.N.
2006, \apjl, 642, L13 

\bibitem[Gawiser \etal(2006b)]{musyc} Gawiser, E., van Dokkum, P.G., 
Herrera, D., Maza, J., Castander, F.J., Infante, L., Lira, P., Quadri, R., 
Toner, R., Treister, E., Urry, C.M., Altmann, M., Assef, R., Christlein, D., 
Coppi, P.S., Dur\'an, M.F., Franx, M., Galaz, G., Huerta, L., Liu, C., 
L\'opez, S., M\'endez, R., Moore, D.C., Rubio, M., Ruiz, M.T., Toft, S., \&
Yi, S.K.  2006, \apjs, 162, 1

\bibitem[Gawiser \etal(2007)]{gawiser07} Gawiser, E., Francke, H., Lai, K., 
Schawinski, K., Gronwall, C., Ciardullo, R., Quadri, R., Orsi, A., 
Barrientos, L.F., Blanc, G.A., Fazio, G., Feldmeier, J.J., Huang, J.,
Infante, L., Lira, P., Padilla, N., Taylor, E.N., Treister, E., Urry, C.M., 
van Dokkum, P.G., \& Virani, S.N. 2007, \apj, 671, 278 

\bibitem[Giavalisco \etal(2004)]{GOODS} Giavalisco, M., \etal\ 2004, \apjl,
600, L93

%\bibitem[Gronwall \etal(2010)]{gronwall10} Gronwall, C., Bond, N.A.,
%Ciardullo, R., Gawiser, E., Altmann, M., Blanc, G.A., \& Feldmeier, J.J.
%2010, arXiv:1005.3006 

\bibitem[Gronwall \etal(2007)]{g+07} Gronwall, C., Ciardullo, R., Hickey, T., 
Gawiser, E., Feldmeier, J.J., van Dokkum, P.G., Urry, C.M., Herrera, D., 
Lehmer, B.D., Infante, L., Orsi, A., Marchesini, D., Blanc, G.A., 
Francke, H., Lira, P., \& Treister, E. 2007, \apj, 667, 79 (Gr07)

\bibitem[Guaita \etal(2011)]{guaita+11} Guaita, L., Acquaviva, V.,
Padilla, N., Gawiser, E., Bond, N.A., Ciardullo, R., Tresiter, E., 
Kurczynski, P., Gronwall, C., Lira, P., \& Schawinski, K. 2011, 
\apj, 733, 114

\bibitem[Guaita \etal(2010)]{guaita+10} Guaita, L., Gawiser, E., Padilla, N., 
Francke, H., Bond, N.A., Gronwall, C., Ciardullo, R., Feldmeier, J.J., 
Sinawa, S., Blanc, G.A., \& Virani, S. 2010, \apj, 714, 255 (Gu10)

\bibitem[Hamuy \etal(1992)]{hamuy92} Hamuy, M., Walker, A.R., 
Suntzeff, N.B., Gigoux, P., Heathcote, S.R., \& Phillips, M.M. 1992, 
\pasp, 104, 533 

\bibitem[Hayes \etal(2010)]{hayes} Hayes, M., \"Ostlin, G., Schaerer, D., 
Mas-Hesse, J.M., Leitherer, C., Atek, H., Kunth, D., Verhamme, A., 
de Barros, S., \& Melinder, J. 2010, \nat, 464, 562 

\bibitem[Hansen \& Oh(2006)]{hansen} Hansen, M., \& Oh, S.P. 2006, 
\mnras, 367, 979 

\bibitem[Hill \etal(2008)]{hetdex} Hill, G.J.,  Gebhardt, K., Komatsu, E., 
Drory, N., MacQueen, P.J., Adams, J., Blanc, G.A., Koehler, R., Rafal, M., 
Roth, M.M., Kelz, A., Gronwall, C., Ciardullo, R., \& Schneider, D.P.
2007, A.S.P. Conference Series 399, Panoramic Views of Galaxy Formation 
and Evolution, ed.~T. Kodama, T. Yamada, \& K. Aoki (San Francisco: 
Astronomical Society of the Pacific), 115

\bibitem[Hogg \etal(1998)]{hogg98} Hogg, D.W., Cohen, J.G., 
Blandford, R., \& Pahre, M.A. 1998, \apj, 504, 622 

\bibitem[Hoversten \etal(2009)]{UVOT} Hoversten, E.A.,  Gronwall, C., 
Vanden Berk, D.E., Koch, T.S., Breeveld, A.A., Curran, P.A., Hinshaw, D.A., 
Marshall, F.E., Roming, P.W.A., Siegel, M.H., \& Still, M.  2009, \apj, 705, 
1462 

\bibitem[Iwata \etal(2007)]{iwata} Iwata, I., Ohta, K., Tamura, N., 
Akiyama, M., Aoki, K., Ando, M., Kiuchi, G., \& Sawicki, M. 2007, 
\mnras, 376, 1557 

\bibitem[Jacoby \etal(1987)]{jqa} Jacoby, G.H., Quigley, R.J., \& Africano,
J.L. 1987, \pasp, 99, 672

\bibitem[Kennicutt(1998)]{kennicutt} Kennicutt, R.C. 1998, \araa, 36, 189

\bibitem[Kobayashi \etal(2010)]{kobayashi+10} Kobayashi, M.A.R., 
Totani, T., \& Nagashima, M. 2010, \apj, 708, 1119 

\bibitem[Le F\`evre \etal(2005)]{vvds} Le F\`evre, O., et al. 2005, 
\aap, 439, 845 

\bibitem[Lehmer \etal(2005)]{lehmer05} Lehmer, B.D., Brandt, W.N., 
Alexander, D.M., Bauer, F.E., Schneider, D.P., Tozzi, P., Bergeron, J., 
Garmire, G.P., Giacconi, R., Gilli, R., Hasinger, G., Hornschemeier, A.E., 
Koekemoer, A.M., Mainieri, V., Miyaji, T., Nonino, M., Rosati, P., 
Silverman, J.D., Szokoly, G., \& Vignali, C. 2005, \apjs, 161, 21

\bibitem[Lehmer \etal(2010)]{lehmer10} Lehmer, B.D., Alexander, D.M., 
Bauer, F.E., Brandt, W.N., Goulding, A.D., Jenkins, L.P., Ptak, A., 
\& Roberts, T.P. 2010, \apj, 724, 559 

\bibitem[Leitherer \etal(1999)]{leitherer99} Leitherer, C., Schaerer, D., 
Goldader, J.D., Gonz\'alez Delgado, R.M., Robert, C., Kune, D.F., 
de Mello, D.F., Devost, D., \& Heckman, T.M.  1999, \apjs, 123, 3 

\bibitem[Luo \etal(2008)]{luo} Luo, B., Bauer, F.E., Brandt, W.N., 
Alexander, D.M., Lehmer, B.D., Schneider, D.P., Brusa, M., Comastri, A., 
Fabian, A.C., Finoguenov, A., Gilli, R., Hasinger, G., Hornschemeier, A.E., 
Koekemoer, A., Mainieri, V., Paolillo, M., Rosati, P., Shemmer, O., 
Silverman, J.D., Smail, I., Steffen, A.T., \& Vignali, C. 2008,
\apjs, 179, 19 

\bibitem[Mao \etal(2007)]{mao+07} Mao, J., Lapi, A., Granato, G.L.,
de Zotti, G., \& Danese, L. 2007, \apj, 667, 655 

\bibitem[Meurer \etal(1999)]{meurer+99} Meurer, G.R., Heckman, T.M.,
\& Calzetti, D. 1999, \apj, 521, 64 

\bibitem[Monet \etal(2003)]{usno-b} Monet, D.G., Levine, S.E., Canzian, B., 
Ables, H.D., Bird, A.R., Dahn, C.C., Guetter, H.H., Harris, H.C., 
Henden, A.A., Leggett, S.K., Levison, H.F., Luginbuhl, C.B., Martini, J., 
Monet, A.K.B., Munn, J.A., Pier, J.R., Rhodes, A.R., Riepe, B., Sell, S., 
Stone, R.C., Vrba, F.J., Walker, R.L., Westerhout, G., Brucato, R.J.,
 Reid, I.N., Schoening, W., Hartley, M., Read, M.A., \& Tritton, S.B. 2003,
\aj, 125, 984 

\bibitem[Neufeld(1991)]{neufeld} Neufeld, D.A. 1991, \apjl, 370, L85 

\bibitem[Nilsson \etal(2009a)]{nilsson09a} Nilsson, K.K., Tapken, C., 
M\"oller, P., Freudling, W., Fynbo, J.P.U., Meisenheimer, K., Laursen, P., 
\"Ostlin, G. 2009a, \aap, 498, 13

\bibitem[Nilsson \etal(2009b)]{nilsson09b} Nilsson, K.K., M\"oller-Nilsson, O., 
M\"oller, P., Fynbo, J.P.U., \& Shapley, A.E. 2009b, \mnras, 400, 232 

\bibitem[Nilsson \etal(2011)]{nilsson11} Nilsson, K.K., 
\"Ostlin, G., M{\o}ller, P., M\"oller-Nilsson, O., Tapken, C., 
Freudling, W., \& Fynbo, J.P.U. 2011, \aap, 529, A9

\bibitem[Ouchi \etal(2008)]{ouchi+08} Ouchi, M.,  Shimasaku, K,, Akiyama, M,, 
Simpson, C., Saito, T., Ueda, Y., Furusawa, H., Sekiguchi, K., Yamada, T., 
Kodama, T., Kashikawa, N., Okamura, S., Iye, M., Takata, T., Yoshida, M., \&
Yoshida, M.  2008, \apjs, 176, 301 

\bibitem[Ouchi \etal(2010)]{ouchi+10} Ouchi, M., Shimasaku, K., Furusawa, H., 
Saito, T., Yoshida, M., Akiyama, M., Ono, Y., Yamada, T., Ota, K., 
Kashikawa, N., Iye, M., Kodama, T., Okamura, S., Simpson, C., \&
Yoshida, M. 2010, \apj, 723, 869 

\bibitem[Persic \etal(2004)]{persic} Persic, M., Rephaeli, Y., Braito, V., 
Cappi, M., Della Ceca, R., Franceschini, A., \& Gruber, D.E. 2004, 
\aap, 419, 849

%\bibitem[Popesso \etal(2009)]{popesso} Popesso, P., Dickinson, M., 
%Nonino, M., Vanzella, E., Daddi, E., Fosbury, R.A.E., Kuntschner, H., 
%Mainieri, V., Cristiani, S., Cesarsky, C., Giavalisco, M., Renzini, A., 
%\& the GOODS Team 2009, \aap, 494, 443 

%\bibitem[Ranalli \etal(2003)]{ranalli} Ranalli, P., Comastri,
%A., \& Setti, G. 2003, \aap, 399, 39

\bibitem[Ravindranath \etal(2006)]{rav06} Ravindranath, S., Giavalisco, M., 
Ferguson, H.C., Conselice, C., Katz, N., Weinberg, M., Lotz, J., 
Dickinson, M., Fall, S.M., Mobasher, B., \& Papovich, C. 2006,
\apj, 652, 963 

\bibitem[Reddy \etal(2008)]{reddy} Reddy, N.A., Steidel, C.C.,
Pettini, M., Adelberger, K.L., Shapley, A.E., Erb, D.K., \& Dickinson, M.
2008, \apjs, 175, 48

\bibitem[Rix \etal(2004)]{GEMS} Rix, H.-W.,  Barden, M., Beckwith, S.V.W.,
Bell, E.F., Borch, A., Caldwell, J.A.R., H\"aussler, B., Jahnke, K.,
Jogee, S., McIntosh, D.H., Meisenheimer, K., Peng, C.Y.,
Sanchez, S.F., Somerville, R.S., Wisotzki, L., \& Wolf, C. 2004,
\apjs, 152, 163

\bibitem[Salpeter(1955)]{salpeter} Salpeter, E.E. 1955, \apj, 121, 161 

\bibitem[Schechter(1976)]{schechter} Schechter, P. 1976, \apj, 203, 297

\bibitem[Schiminovich \etal(2003)]{Galex} Schiminovich, D., Arnouts, S., 
Milliard, B., \& the GALEX Science Team 2003, BAAS, 35, 1371 

\bibitem[Scholz \& Stephens(1987)]{adtest} Scholz, F.W., \& Stephens, M.A.
1987, J.~Amer.~Stat.~Assoc., 82, 918

\bibitem[Shapley \etal(2003)]{shapley} Shapley, A.E., Steidel, C.C., 
Pettini, M., \& Adelberger, K.L. 2003, \apj, 588, 65 

\bibitem[Shioya \etal(2009)]{shioya} Shioya, Y., Taniguchi, Y., 
Sasaki, S.S., Nagao, T., Murayama, T., Saito, T., Ideue, Y., Nakajima, A., 
Matsuoka, K., Trump, J., Scoville, N.Z., Sanders, D.B., Mobasher, B., 
Aussel, H., Capak, P., Kartaltepe, J., Koekemoer, A., Carilli, C., 
Ellis, R.S., Garilli, B., Giavalisco, M., Kitzbichler, M.G., Impey, C.,
LeFevre, O., Schinnerer, E., \& Smolcic, V. 2009, \apj, 696, 546 

\bibitem[Stone(1977)]{stone} Stone, R.P.S. 1977, \apj, 218, 767

\bibitem[Strateva \etal(2005)]{strateva} Strateva, I.V., Brandt, W.N., 
Schneider, D.P., Vanden Berk, D.G., \& Vignali, C. 2005, \aj, 130, 387 

\bibitem[Szokoly \etal(2004)]{szokoly} Szokoly, G.P., Bergeron, J., 
Hasinger, G., Lehmann, I., Kewley, L., Mainieri, V., Nonino, M., Rosati, P., 
Giacconi, R., Gilli, R., Gilmozzi, R., Norman, C., Romaniello, M., 
Schreier, E., Tozzi, P., Wang, J.X., Zheng, W., \& Zirm, A. 2004, 
\apjs, 155, 271 

\bibitem[Vanzella \etal(2008)]{vanzella+08} Vanzella, E., Cristiani, S., 
Dickinson, M., Giavalisco, M., Kuntschner, H., Haase, J., Nonino, M., 
Rosati, P., Cesarsky, C., Ferguson, H.C., Fosbury, R.A.E., Grazian, A., 
Moustakas, L.A., Rettura, A., Popesso, P., Renzini, A., Stern, D., \& the
GOODS Team 2008, \aap, 478, 83 

\bibitem[Vanzella \etal(2009)]{vanzella+09} Vanzella, E., 
Giavalisco, M., Dickinson, M., Cristiani, S., Nonino, M., Kuntschner, H., 
Popesso, P., Rosati, P., Renzini, A., Stern, D., Cesarsky, C., 
Ferguson, H.C., \& Fosbury, R.A.E. 2009, \apj, 695, 1163 

\bibitem[Virani \etal(2006)]{virani} Virani, S.N., Treister, E., 
Urry, C.M., \& Gawiser, E. 2006, \aj, 131, 2373 

\bibitem[Wolf \etal(2004)]{combo17} Wolf, C.,  Meisenheimer, K., 
Kleinheinrich, M., Borch, A., Dye, S., Gray, M., Wisotzki, L., Bell, E.F., 
Rix, H.-W., Cimatti, A., Hasinger, G., \& Szokoly, G. 2004, \aap, 421, 913

\end{thebibliography}
\end{document}